\documentclass[preprint,12pt,longtitle,authoryear]{elsarticle}
\usepackage{amsmath}
\usepackage{amssymb}
\usepackage{amsthm}
\usepackage{amstext}
\usepackage{amsfonts}
\usepackage{setspace}
\usepackage{fixltx2e}
\usepackage{textcomp}
\usepackage{verbatim}
\usepackage[english]{babel}
\usepackage{pifont}
\usepackage{color}
\usepackage{lscape}
\usepackage[normalem]{ulem}
\usepackage{booktabs}
\usepackage{float}
\usepackage{latexsym}
\usepackage{epsfig}
\usepackage{graphicx}
\usepackage{bm}
\usepackage{array}
\usepackage{ifthen}
\usepackage[ansinew]{inputenc}
\usepackage{subfigure}

\doublespace

\newcommand {\debeq}	{\begin{eqnarray*}}
\newcommand {\fineq}	{\end{eqnarray*}}

\newcommand{\bE}{{\mathbb E}}
\newcommand{\bP}{{\mathbb P}}
\newcommand{\cT}{{\mathcal T}}

\newcommand	{\PP}{\mathbb{P}}

\newtheorem	{thm}		{Theorem}[section]

\newtheorem	{lem} 	[thm]	{Lemma}

\newtheorem	{prop}	[thm]{Proposition}
\newtheorem	{cor}		[thm]{Corollary}

\newcommand	{\indic}	[1]
{{\bf{1}}_{\{#1\}}}

\newenvironment{myindentpar}[1]%
     {\begin{list}{}%
             {\setlength{\leftmargin}{#1}}%
             \item[]%
     }
     {\end{list}}

\title{Predicting the loss of phylogenetic diversity under non-stationary diversification models}

\author[cnrs]{Amaury Lambert}
\ead{amaury.lambert@upmc.fr}

\author[uc]{Mike Steel\corref{cor}}
\ead{mike.steel@canterbury.ac.nz}

\cortext[cor]{Corresponding author. Email: {\tt mike.steel@canterbury.ac.nz}, phone: +6421329705.}
\address[cnrs]{Collège de France, Center for Interdisciplinary Research in Biology CNRS UMR 7241, Paris, France}
\address[uc]{Biomathematics Research Centre, University of Canterbury, Private Bag 4800, Christchurch, New Zealand}

\begin{document}

\begin{abstract}
For many taxa, the current high rates of extinction are likely to result in a significant loss of biodiversity. 
The evolutionary heritage of biodiversity  is frequently quantified by a measure called phylogenetic diversity (PD). We predict the loss of PD under a wide class of  phylogenetic tree models, where speciation rates and extinction rates may be time-dependent,  and assuming independent random species extinctions at the present. We study the loss of PD when $K$ contemporary species are selected uniformly at random from the $N$ extant species as the surviving taxa, while the remaining $N-K$ become extinct. We consider two models of species sampling, the so-called  field of bullets model, where each species independently survives the extinction event at the present with probability $p$, and a model for which the number of surviving species is fixed.  
 
 We provide explicit formulae for the expected remaining PD in both models, conditional on $N=n$, conditional on $K=k$, or conditional on both events. When $N=n$ is fixed, we show the convergence  to an explicit deterministic limit of the ratio of new to initial PD, as $n\to\infty$, both under the field of bullets model, and when $K=k_n$ is fixed and depends on $n$ in such a way that $k_n/n$ converges to $p$. We also prove the convergence of this ratio as $T\to\infty$ in the supercritical, time-homogeneous case, where $N$ simultaneously goes to $\infty$, thereby strengthening previous results of Mooers {\em et al.} (2012).\\
\end{abstract}


\maketitle

\section{Introduction}
\subsection{Phylogenetic diversity}

A typical question arising in  biodiversity conservation is the following:

    \begin{myindentpar}{1cm}
 ``If  10\% of taxa from some clade were to randomly disappear in the next 100 years due to current high rates of extinction,  how much evolutionary heritage would be lost?''
    \end{myindentpar}
    
The answer depends on many factors, the first of which is how one measures evolutionary heritage. Here, we adopt phylogenetic diversity (PD) for this purpose -- it assigns to any (surviving) subset of taxa the sum of the branch lengths of the evolutionary tree that span those taxa and the root of the tree \citep{fai}. Thus, one can consider the ratio of the PD after a rapid mass extinction event (the `surviving PD' score) to the initial  PD score as a measure of  the  relative PD loss.

A second important factor in answering this question is the interplay of the tree shape and the process of extinction at the tips.  
For example, the extinction of a taxon at the end of a long pendant edge of the tree will lead to greater PD loss than the extinction of a taxon on a short pendant edge.  However, this is just part of the story, as interaction effects also occur -- for instance, the extinction of two closely related taxa on short pendant edges that form a cherry in the tree at the end of a very long interior edge will lead to far more PD loss than the extinction of two taxa with short to moderate pendant edge lengths that do not form a cherry.

This interplay of tree shape and possible taxon extinction scenarios will vary from data-set to data-set, and will generally depend on a large number of parameters (related to the tree, its branch lengths, the extinction risks of different taxa and how they are correlated), some of which are often not known with any precision. 

In this paper, we establish general results and properties concerning relative PD loss,  by showing how it can be estimated by closed-form formulae based on
 stochastic diversification models that describe how phylogenetic trees arise under speciation and extinction models \citep{ald, ald3, mor, pur, rab}, together with a simple  `field of bullets' model of random, instantaneous extinctions at the present.  Thus, our approach is in a similar spirit to \citet{nee} and the more recent paper by \citet{moo}, but our results generalize and strengthen these earlier results in some important ways:

\begin{itemize}
\item Most of our results  allow the speciation rate $b(t)$ to depend on time $t$, and the extinction rate $d(t;x)$ to depend on time $t$ and/or on a non-heritable trait $x$ (i.e. a discrete or continuous trait changing in the same way in all species, for example the age of the species); this generalizes the classical (constant rate) birth--death model where $b(t)=b$ and $d(t;x)=d$ for constants $b,d \geq 0$, thereby allowing greater biological realism.
\item Rather than studying the limiting ratio of expected surviving PD to expected initial PD  (as in \citet{moo}), we analyse the actual ratio of new to initial PD and establish its convergence to  explicitly computable functions under each of two limiting processes (increasing number of taxa and increasing time). This provides for statements with greater statistical precision. 
\item We also present explicit exact formulae for the expected surviving PD (and the expected loss of PD), given a fixed initial number of taxa and the depth of the tree under  sudden random mass extinctions at the present. We also provide a formula for when we explicitly condition on the number of taxa that survive this sudden extinction event.

\end{itemize}
Before proceeding to describe our results, we summarize some standard terms in probability theory that will be used throughout this paper.

\subsection{Coalescent point processes}

We model species diversification by a binary branching process, where species are viewed as autonomous and independent particles that speciate and become extinct at random times. We assume that the diversification process starts with one species at time 0, that species speciate during their lifetime at some rate $b(t)$ that may depend on absolute time $t$, and become extinct at some rate $d(t;x)$ that may depend both on absolute time $t$ and on the age of the species or any non-heritable trait $x$ varying with the same probability transitions for all species. When $x$ is the age, the tree embedded in continuous time thus generated is called a (time-inhomogeneous) splitting tree \citep{gei, lam1}. 

It has been known since \citet{lam1} and \citet{lam3}, that, conditional on survival up until time $T$, 
 for this general class of diversification processes, the \emph{reconstructed tree} seen at $T$, i.e. the tree spanned by all species extant at time $T$ is a \emph{coalescent point process}. This means that:
\begin{enumerate}
\item the number $N_T$ of species extant at time $T$ is geometric with a success probability of, say, $a_T$;
\item conditional on $N_T=n$, all node depths are i.i.d.;  
\item the shape of the tree is uniform among (oriented) ranked tree shapes.
\end{enumerate}
We denote by $A_T$ (or simply $A$) a random variable  having the common distribution of these node depths. There is a random variable $H$ such that $\bP(H\ge T)=a_T$ (sometimes simply denoted $a$) and $A_T$ is distributed as $H$ conditional on $H\le T$. It will be convenient to define the {\em scale function} $W$ by:
\begin{equation}
\label{Weq}
W(t) := 1/\bP(H\ge t).
\end{equation}
We will also write $F_T(t):=\bP(A_T<t)$ and $F(t)= \bP(H<t)$, so that $W= 1/(1-F)$  and
\begin{equation}
\label{eqn:W}
a_T=\bP(H\ge T) = 1-F(T)\quad \mbox{ and }\quad F_T(t) := \bP(A_T<t) = \bP(H<t \mid H<T) = \frac{F(t)}{F(T)} .
\end{equation}
Note that the law of a coalecent point process is totally characterised from the knowledge of the function $W$.

\subsection{Links to macroevolutionary models of diversification}
Here, we provide means of computing the function $W$ characterising the reconstructed tree in terms of the parameters of the underlying macroevolutionary model of diversification, i.e., the speciation rate $b(t)$  and the extinction rate $d(t;x)$ of species carrying trait value $x$ at time $t$. The results stated here can be found in \citet{lam3}. This subsection can be skipped in a primary reading.

In the case when the diversification rates do \emph{not} depend on a trait (Markovian case), but may depend on time, $W$ is explicit. Setting $r(t):= b(t)-d(t)$, we have the following expression
\begin{equation}
\label{Weq-inhomogeneous}
W(t) = 1+ \int_0^t  b(T-s)\, e^{\int_{T-s}^T  r(u)\,du}\,ds\qquad t\in[0,T].
\end{equation}
In all other cases, there is generally no explicit expression for $W$. However, if one knows the density at time $s$ of the lifetime of a species born at time $t$, say $h(t,s)$, then $W$ is the unique solution to 
\begin{equation}
\label{eqn diff}
W'(t) = b(T-t)\,\left( W(t) -  \int_0^t W(s)\, h(T-t, T-s)\, ds\right)\qquad  t\in[0,T],
\end{equation}
satisfying $W(0)=1$.
Furthermore, the density $h$ can be computed in many cases of biological interest. 
For example, if the trait is the age, then
\begin{equation}
\label{eqn:h age}
h(t,s) = d(s, s-t)\,e^{-\int_t^s  d(u, u-t)\,du}
\end{equation}
In the time-homogeneous case, that is, when (a) the diversification rates do not depend on time; (b) the initial value of the trait of a new species (can be random but) does not depend on the speciation time; (c) the probability transitions of the trait dynamics do not depend on time either, then all species have equally distributed lifetimes. The common density $g$ of these lifetimes satisfies $h(t,s) = g(s-t)$, and so \eqref{eqn diff} becomes
$$
W'(t) = b\,\left( W(t) -  \int_0^t W(s)\, g(t-s)\, ds\right) \qquad t \ge 0.
$$
In this case, the function $W$ does \emph{not} depend on $T$ and is the unique non-negative solution to
\begin{equation}
\label{WeqLaplace}
\int_0^\infty W(t)\,e^{-xt}\, dt = \left( x- b + b \int_0^\infty g(u) \,e^{-xu}\, du\right)^{-1}\qquad x \ge 0.
\end{equation}
If in addition to time-homogeneity we assume that the trait is the age, then by \eqref{eqn:h age}, we get the following expression for $g$
\begin{equation}
\label{eqn:lifetime density}
g(a) = d(a)\,e^{-\int_0^a d(s)}\,ds\qquad a\ge 0.
\end{equation}
The intersection between the Markovian case and the time-homogeneous case is the linear birth--death process, with a constant speciation rate $b$ and a constant extinction rate $d$. Then $g$ is the exponential density with parameter $d$ and it can easily be verified that the common solution to  \eqref{Weq-inhomogeneous} and \eqref{WeqLaplace} is given by:
\begin{equation}
\label{Weqcritical}
W(t) = 1+bt,
\end{equation}
in the critical case when $b=d$, whereas when $r:=b-d \not =0$,
\begin{equation}
\label{Weqnotcritical}
W(t)= 1+\frac{b}{r}(e^{rt}-1).
\end{equation}
In  conclusion, everything that follows holds under a general lineage-based branching model with speciation rate that possibly depends on time and an extinction rate that is possibly dependent on a non-heritable trait and time, starting with one single species and conditioned to survive to time $T$. Practically speaking, the expressions in \eqref{Weq} and \eqref{eqn:W} can be: (a) expressed explicitly in the case when rates do not depend on a trait, thanks to Equation \eqref{Weq-inhomogeneous}; (b) evaluated numerically when the extinction rate further depends on age, thanks to Equations \eqref{eqn diff} and \eqref{eqn:h age}; evaluated numerically when the extinction rate further depends on a non-heritable trait, provided the probability density $h$ of lifetimes is known (see \citet{lam3}, for additional formulae allowing the treatment of this general case).

\subsection{Terminology from probability theory}
Recall first that a  {\em Bernoulli random variable} has just two outcomes (0,1), with 1 referred to as a `success'.   Given a sequence of  independent and identically distributed (i.i.d.) Bernoulli random variables $X_1, X_2, \ldots$, where $X_i$ has success probability $p$, the random variable $J$ that specifies the first value $j\geq 1$ for which
$X_j=1$ is a (shifted)  {\em geometric random variable}  (with success probability $p$); its distribution is easily seen to be $\PP(J=j) = (1-p)^{j-1} p$, for $j=1,2,\ldots$.
For example, the number of rolls of a fair die until the number 4 first appears is a  geometric random variable with success probability $p=1/6$.

A sequence of random variables $X_n$ {\em converges in probability} to some constant value $c$ if the probability that $X_n$ differs from $c$ by every given positive value 
$\epsilon$ tends to zero as $n \rightarrow \infty$. For example the proportion of tosses of a fair coin that result in a head converges in probability to $c=1/2$ (by the weak law of large numbers, or by the
central limit theorem).
A stronger notion is to say that $X_n$ converges {\em almost surely} to some constant $c$, which means that any realization (e.g. numerical simulation) of the sequence $(X_n)$ converges to $c$ with probability 1.   This actually holds also for the coin-tossing
example (by the strong law of large numbers).  Almost sure convergence implies convergence in probability but the converse need not hold.

\subsection{Summary of results}

We use the coalescent point process characterization of the reconstructed tree to study the loss of phylogenetic diversity when contemporary species are randomly removed from the standing species set. For any time-calibrated phylogeny, the \emph{phylogenetic diversity} (PD) is the total sum of branch lengths, also called the total length of the tree. We consider two models of random species removal. 

The first model, called the `field of bullets' (\citet{rau}, \citet{nee}, \citet{pur}), makes the assumption that every contemporary species, (i.e. every tip of the phylogeny) is independently removed with probability $1-p$, where $p$ will be called the \emph{sampling probability}. We will denote by $K$ the total number of sampled and so retained species, by $S_N(p)$ the remaining PD after the passage of the field of bullets, and by $K_n$ and $S_n(p)$, respectively, the same quantities when conditioning the initial number $N$ of species (i.e. before sampling) to equal $n$.

\begin{figure}[htb]
\centering
\includegraphics[scale=0.8]{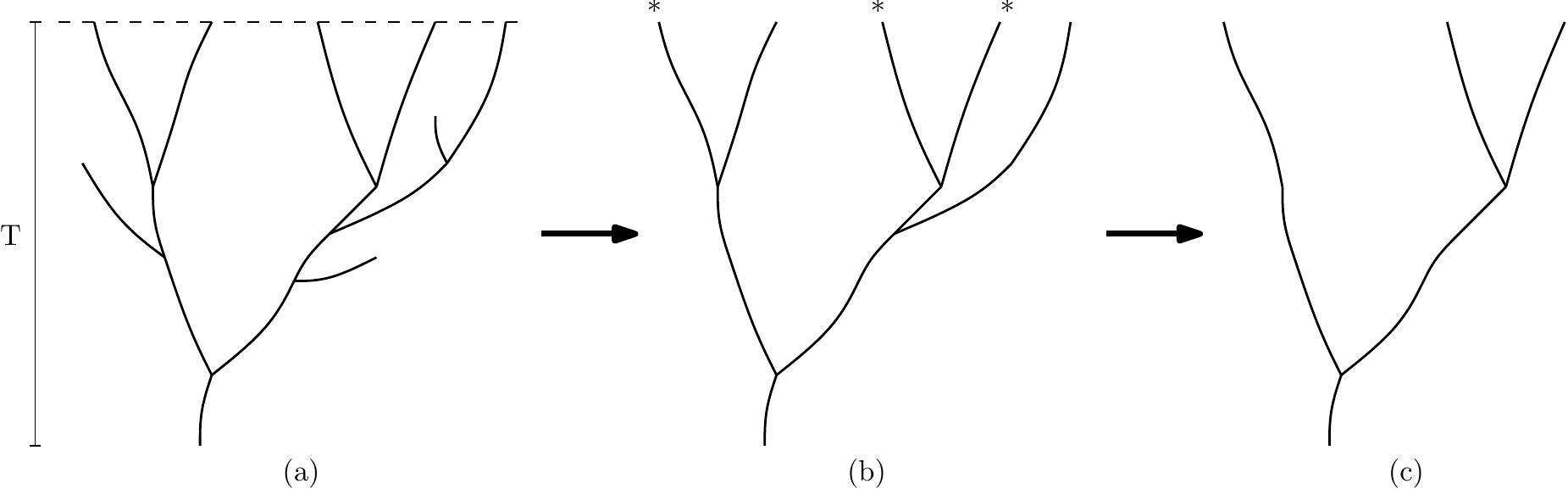}
\caption{(a) An evolutionary tree as it arises under a continuous model of speciation and extinction, observed at the present time $T$.  (b) The reconstructed tree  obtained by deleting lineages that do not survive to the present. The leaves are now subject to a further extinction event at the present (e.g. under a `field of bullets' model ). The tree connecting the surviving leaves (indicated by * in (b)) is shown in (c).}
\label{overview_fig}
\end{figure}

The second model consists in fixing the number of sampled species, to $k$ say, and to sample these $k$ species uniformly at random as soon as $N\ge k$. We will denote by $S_{N,k}$ the remaining PD after sampling these $k$ species (and removing all others); similarly, $S_{n,k}$ denotes the same quantity when conditioning on the initial number of species to equal $n$. 

Fig.~\ref{overview_fig} provides an overview of the three steps in the processes we consider: (a) the phylogenetic tree generated by a speciation  and extinction model; (b) the reconstructed tree; and (c) the  tree that connects leaves that survive mass extinction (e.g. a field of bullets model or sampling $k$ species from $n$).

We will also use the notation $S_N=S_{N,N}= S_N(1)$ for the initial PD. It will be convenient to add an exponent $\star$ to denote the phylogenetic diversity minus the stem age $T$, e.g., $S_{n,k}^\star := S_{n,k}-T$ for $k>0$.  Also notice that, conditional on $K_n=k$ under the field of bullets model, $S_n(p)$ does not depend on $p$ and is equal to $S_{n,k}$. Conversely, $S_n(p)$ can be seen as a mixture over $k$ of the random values $S_{n,k}$, where the mixing distribution is the binomial distribution with parameters $n$ and $p$.\\

In this paper, we characterize the distribution of the quantities $S_N(p)$, $S_n(p)$, $S_{N,k}$ and $S_{n,k}$, and we provide explicit formulae for their expectations. In Theorem \ref{thm1}, we prove the convergence in probability, to the same explicit deterministic limit $\pi_T$, of $S_n(p)/S_n$ as $n\to\infty$ and $S_{n,k_n}/S_{n}$ as $n\to\infty$ and $k_n/n\to p$. It can be expressed as $p\bE(Y_T)/\bE(A_T)$, where:
\begin{itemize}
\item  $Y_T$  is the maximum of $G$ independent copies of $A_T$; and 
\item  $G$ is a  geometric random variable with success probability $p$. 
\end{itemize}
Namely,
$$
\lim_{n\to\infty}\frac{S_n(p)}{S_n}=\lim_{n\to\infty}\frac{S_{n,k_n}}{S_n} =:\pi_T(p),
$$
where the first convergence holds almost surely and the second one holds in probability (with $k_n/n\to p$), and
\begin{equation}   
\label{eqn:main1}
\pi_T(p) = \frac{p\bE(Y_T)}{\bE(A_T)} = \frac{p\int_0^T \frac{1-F_T(t)}{1-(1-p)F_T(t)}\, dt}{ \int_0^T (1-F_T(t))\, dt}.
\end{equation}
Notice that the expectations of these ratios, as well as the ratios of the expectations $\bE(S_n(p))/\bE(S_n)$ and $\bE(S_{n,k_n})/\bE(S_n)$, also converge to $\pi_T(p)$.\\

In Theorem \ref{thm2}, in the case of time-homogeneous supercritical branching processes (where $F$ does not depend on $T$ and $N_T\to\infty$ as $T\to\infty$ conditional on                                     survival), we write $S_T(p)$ instead of $S_N(p)$ and we prove the convergence of $S_T(p)/S_T$ to a deterministic limit denoted $\pi_\infty(p)$, as time $T\to\infty$.  It can be expressed as  $p\bE(Z)/\bE(H)$, where
\begin{itemize}
\item $Z$ is the maximum of $G$ independent copies of $H$. 
\end{itemize}
Namely,
$$
\lim_{T\to\infty}\frac{S_T(p)}{S_T}=:\pi_\infty(p),
$$
where the last convergence holds in probability and
\begin{equation}
\label{eqn:main2}
\pi_\infty(p) = \frac{p\bE(Z)}{\bE(H)} = \frac{p\int_0^\infty \frac{1-F(t)}{1-(1-p)F(t)}\, dt}{ \int_0^\infty (1-F(t))\, dt}.
\end{equation}
Notice that the ratio $\bE(S_T(p))/\bE(S_T)$ also converges to $\pi_\infty(p)$, and note also the similarity between \eqref{eqn:main1} and \eqref{eqn:main2}. Since $F_T(t)=F(t)/F(T)$, we have the convergence of $\pi_T(p)$ to $\pi_\infty(p)$ as $T\to +\infty$ for any time-homogeneous supercritical branching processes.

In the case of birth--death processes with a constant speciation rate $b$, a constant extinction rate $d$ and diversification rate $r=b-d$, we show in Corollary \ref{coro} that
$$
\pi_\infty(p)=
\begin{cases}
\frac{dp}{bp-r}\frac{\ln(bp/r)}{\ln (b/r)}, & \mbox{ if }  b>r\not= bp;\\
 -\frac{p\ln (p)}{1-p}, &  \mbox{ if }  b=r> bp;\\
-\frac{1-p}{\ln (p)}, & \mbox{ if }  b>r= bp.  
\end{cases}
$$
The  convergence of the ratio of expectations $\bE(S_T(p))/\bE(S_T(1))$ to $\pi_\infty(p)$ in the case of birth--death trees was first displayed in \citet{moo} (equation (7) of that paper).\\

Before stating these convergence results, we provide exact expressions for the expected surviving PD, $\bE(S_n(p))$, when conditioning on the number $n$ of initial taxa and the depth of the tree, $n$ being fixed and finite. From this, one can immediately derive exact expressions for the expected {\em loss} of PD, which is $\bE(S_n(1) - S_n(p))$, as well as for the ratio of expected new-to-initial PD (i.e.  $\bE(S_n(p)) /\bE(S_n(1))$). Moreover, if we further condition on the number $k$ of surviving taxa (after the passage of the field of bullets), one can also provide exact expressions for  these quantities.

We stress that our results hold for any macroevolutionary model of diversification with no diversity-dependence, (possibly) time-dependent speciation rate and (possibly) time-dependent and/or trait-dependent extinction rate. Thanks to Equations \eqref{Weq-inhomogeneous}, \eqref{Weqcritical} and \eqref{Weqnotcritical}, these results are totally explicit in the case when the rates are only time-dependent or simply constant. They are semi-explicit thanks to \eqref{eqn diff} and \eqref{eqn:h age} in the case when the extinction rate can additionally be age-dependent. Compared to the limits obtained for expected loss of PD in \citet{moo}, the results we obtain here are stronger, because (i) they apply to a wider class of random trees, (ii) the expected loss of PD is given for a wider spectrum of mass extinction models, and  (iii) the convergences are almost sure convergences of new-to-initial PD ratios instead of the convergence of the ratio of their expectations.

We also insist that since phylogenetic diversity is independent of tree topology, and because sampling schemes are also assumed to be independent of tree topology, our results are also valid for any tree obtained from our trees by changing their topology  but keeping the same node depths (e.g.,  for a comb with the same node depths).

\subsection{Why assume a constant value of $p$ across taxa?}

In the  field of bullets model,  each leaf of the reconstructed tree has the same survival probability $p$.  However, often it is clear that different taxa will have higher or lower extinction risks than others, so it would seem that a more realistic extension of this model would allow each leaf $x$ of the reconstructed tree  to have its own survival probability $p(x)$.  This `generalized field of bullets' model \citep{fal} is relevant if we are given a phylogenetic tree  and some indication of taxon survival probabilities  (estimated, for example, from the International Union for Conservation of Nature (IUCN) red list).    However, when the trees are randomly generated, as is the case in the birth--death models we consider here, the identity of the taxa among the leaves is effectively randomized (i.e. each permutation of leaf labels results in a tree having the same probability as the original) so the closest and biologically most realistic analogue of a generalized field of bullets model in this setting would be the following:
 \begin{myindentpar}{1cm}
`Each leaf in the reconstructed tree is independently assigned a survival probability $s$ that is drawn from a fixed probability distribution $G(s)$ on $[0,1]$.'
  \end{myindentpar}
This model is, in fact, stochastically equivalent to the simpler field of bullets mode in which all species have
survival probability $p$, where $p$ is the mean of the distribution $G(s)$. This is due to the observation that any sequence of independent Bernoulli random variables 
in which the success probability $s$ for each variable is drawn independently from a common distribution has the same joint probability distribution as a sequence of i.i.d. Bernoulli variables with a success probability $p$ equal to the mean of that distribution.  In summary, the generalized field of bullets is an important extension on a given phylogenetic tree, if one has good estimates of the extinction risk of particular taxa (e.g., from the IUCN red list), but in the setting of this paper, it provides no additional complication.

\section{Coalescent point processes}

In this section, we define coalescent point processes and review their main properties.  We then explain how these processes can model phylogenetic trees exactly under a wide range of diversification models. 

An \emph{oriented} tree is a rooted, binary tree embedded in the plane where time flows upwards and where mother and daughter are distinguished by putting the daughter to the right of her mother. An \emph{ultrametric} tree is a rooted tree whose tip points are all at the same graph distance to the root point.

A \emph{coalescent point process} is a random, oriented, ultrametric tree with edge lengths, where the tips are numbered $0,1,2,\ldots$ from left to right, starting with a single root point, and which satisfies the independence property stated below. 

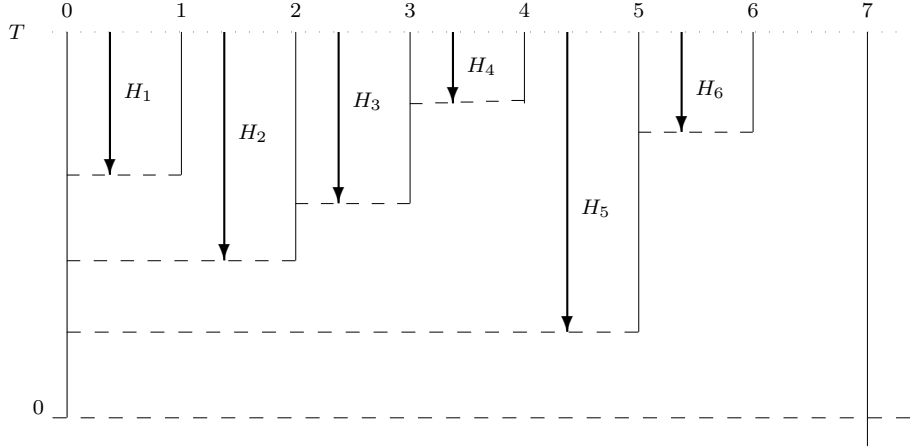
\begin{figure}[ht]
\unitlength 1.9mm 
\linethickness{0.2pt}
\ifx\plotpoint\undefined\newsavebox{\plotpoint}\fi 
\begin{picture}(69,31.5)(-4,3)
\put(4.982,5.982){\line(1,0){.9836}}
\put(6.949,5.982){\line(1,0){.9836}}
\put(8.916,5.982){\line(1,0){.9836}}
\put(10.883,5.982){\line(1,0){.9836}}
\put(12.85,5.982){\line(1,0){.9836}}
\put(14.818,5.982){\line(1,0){.9836}}
\put(16.785,5.982){\line(1,0){.9836}}
\put(18.752,5.982){\line(1,0){.9836}}
\put(20.719,5.982){\line(1,0){.9836}}
\put(22.686,5.982){\line(1,0){.9836}}
\put(24.654,5.982){\line(1,0){.9836}}
\put(26.621,5.982){\line(1,0){.9836}}
\put(28.588,5.982){\line(1,0){.9836}}
\put(30.555,5.982){\line(1,0){.9836}}
\put(32.522,5.982){\line(1,0){.9836}}
\put(34.49,5.982){\line(1,0){.9836}}
\put(36.457,5.982){\line(1,0){.9836}}
\put(38.424,5.982){\line(1,0){.9836}}
\put(40.391,5.982){\line(1,0){.9836}}
\put(42.359,5.982){\line(1,0){.9836}}
\put(44.326,5.982){\line(1,0){.9836}}
\put(46.293,5.982){\line(1,0){.9836}}
\put(48.26,5.982){\line(1,0){.9836}}
\put(50.227,5.982){\line(1,0){.9836}}
\put(52.195,5.982){\line(1,0){.9836}}
\put(54.162,5.982){\line(1,0){.9836}}
\put(56.129,5.982){\line(1,0){.9836}}
\put(58.096,5.982){\line(1,0){.9836}}
\put(60.063,5.982){\line(1,0){.9836}}
\put(62.031,5.982){\line(1,0){.9836}}
\put(63.998,5.982){\line(1,0){.9836}}
\multiput(4.982,32.982)(.983333,0){61}{{\rule{.2pt}{.2pt}}}
\put(2.5,33){\makebox(0,0)[cc]{\scriptsize $T$}}
\put(5.982,22.982){\line(1,0){.8889}}
\put(7.759,22.982){\line(1,0){.8889}}
\put(9.537,22.982){\line(1,0){.8889}}
\put(11.315,22.982){\line(1,0){.8889}}
\put(13.093,22.982){\line(1,0){.8889}}
\put(21.982,20.982){\line(1,0){.8889}}
\put(23.759,20.982){\line(1,0){.8889}}
\put(25.537,20.982){\line(1,0){.8889}}
\put(27.315,20.982){\line(1,0){.8889}}
\put(29.093,20.982){\line(1,0){.8889}}
\put(29.982,27.982){\line(1,0){.8889}}
\put(31.759,28.037){\line(1,0){.8889}}
\put(33.537,28.093){\line(1,0){.8889}}
\put(35.315,28.148){\line(1,0){.8889}}
\put(37.093,28.204){\line(1,0){.8889}}
\put(45.982,25.982){\line(1,0){.8889}}
\put(47.759,25.982){\line(1,0){.8889}}
\put(49.537,25.982){\line(1,0){.8889}}
\put(51.315,25.982){\line(1,0){.8889}}
\put(53.093,25.982){\line(1,0){.8889}}
\put(45.982,11.982){\line(-1,0){.9756}}
\put(44.03,11.982){\line(-1,0){.9756}}
\put(42.079,11.982){\line(-1,0){.9756}}
\put(40.128,11.982){\line(-1,0){.9756}}
\put(38.177,11.982){\line(-1,0){.9756}}
\put(36.225,11.982){\line(-1,0){.9756}}
\put(34.274,11.982){\line(-1,0){.9756}}
\put(32.323,11.982){\line(-1,0){.9756}}
\put(30.372,11.982){\line(-1,0){.9756}}
\put(28.421,11.982){\line(-1,0){.9756}}
\put(26.469,11.982){\line(-1,0){.9756}}
\put(24.518,11.982){\line(-1,0){.9756}}
\put(22.567,11.982){\line(-1,0){.9756}}
\put(20.616,11.982){\line(-1,0){.9756}}
\put(18.664,11.982){\line(-1,0){.9756}}
\put(16.713,11.982){\line(-1,0){.9756}}
\put(14.762,11.982){\line(-1,0){.9756}}
\put(12.811,11.982){\line(-1,0){.9756}}
\put(10.86,11.982){\line(-1,0){.9756}}
\put(8.908,11.982){\line(-1,0){.9756}}
\put(6.957,11.982){\line(-1,0){.9756}}
\put(21.982,16.982){\line(-1,0){.9412}}
\put(20.099,16.982){\line(-1,0){.9412}}
\put(18.217,16.982){\line(-1,0){.9412}}
\put(16.334,16.982){\line(-1,0){.9412}}
\put(14.452,16.982){\line(-1,0){.9412}}
\put(12.57,16.982){\line(-1,0){.9412}}
\put(10.687,16.982){\line(-1,0){.9412}}
\put(8.805,16.982){\line(-1,0){.9412}}
\put(6.923,16.982){\line(-1,0){.9412}}
\put(14,23){\line(0,1){10}}
\put(22,17){\line(0,1){16}}
\put(30,21){\line(0,1){12}}
\put(38,28){\line(0,1){5}}
\put(46,12){\line(0,1){21}}
\put(54,26){\line(0,1){7}}
\put(62,33){\line(0,-1){29}}
\put(14,34.5){\makebox(0,0)[cc]{\scriptsize 1}}
\put(6,34.5){\makebox(0,0)[cc]{\scriptsize 0}}
\put(22,34.5){\makebox(0,0)[cc]{\scriptsize 2}}
\put(30,34.5){\makebox(0,0)[cc]{\scriptsize 3}}
\put(38,34.5){\makebox(0,0)[cc]{\scriptsize 4}}
\put(46,34.5){\makebox(0,0)[cc]{\scriptsize 5}}
\put(54,34.5){\makebox(0,0)[cc]{\scriptsize 6}}
\put(62,34.5){\makebox(0,0)[cc]{\scriptsize 7}}
\thicklines
\put(9,33){\vector(0,-1){10}}
\put(17,33){\vector(0,-1){16}}
\put(25,33){\vector(0,-1){12}}
\put(33,33){\vector(0,-1){5}}
\put(41,33){\vector(0,-1){21}}
\put(49,33){\vector(0,-1){7}}
\put(11,28.75){\makebox(0,0)[cc]{\scriptsize $H_1$}}
\put(19,25.875){\makebox(0,0)[cc]{\scriptsize $H_2$}}
\put(27,28.125){\makebox(0,0)[cc]{\scriptsize $H_3$}}
\put(35,30.625){\makebox(0,0)[cc]{\scriptsize $H_4$}}
\put(43,20.625){\makebox(0,0)[cc]{\scriptsize $H_5$}}
\put(51,29){\makebox(0,0)[cc]{\scriptsize $H_6$}}
\thinlines
\put(6,6){\line(0,1){27}}
\put(4,6.75){\makebox(0,0)[cc]{\scriptsize $0$}}
\end{picture}
\caption{Illustration of a coalescent point process showing the node depths $H_1,\ldots, H_6$ for each of the six consecutive pairs of tips. The node depth $H_7$ is the first one which is larger than $T$.
}
\label{fig : treecoal}
\end{figure}

We call $T$ the stem age of this tree, that is, the common graph distance of the tips to the root point. Note that thanks to the orientation of the tree, if $C_{i,i+k}$ denotes the time elapsed since the lineages of tips $i$ and $i+k$ have diverged, then
\begin{equation}
\label{eqn : def coal}
C_{i,i+k}=\max\{H_{i+1},\ldots,H_{i+k}\}, 
\end{equation}
where $H_i:=C_{i-1,i}$. 
In particular, the genealogical structure is entirely given by the knowledge of the sequence $H_1, H_2,\ldots$, which we will call either \emph{coalescence times} or \emph{node depths} (see Fig.~\ref{fig : treecoal})

\paragraph{Independence property.} There is a random variable  $H$ (whose probability distribution may depend on $T$) such that node depths form a sequence of \textbf{independent, identically distributed random variables, all distributed as $H$,} which terminates at its first value that is larger than $T$.\\

In other words, the number $N_T$ of tips (more simply denoted $N$) in the coalescent point process follows the  geometric distribution with success probability $a_T:=\bP(H\ge T)$, more simply denoted $a$, and, conditional on $N=n$, the node depths $H_1,\ldots, H_n$ are independent copies of $H$ conditioned on $H\le T$. 

From now on, to simplify the notation, we will let $A_T$, or simply $A$, denote a random variable  distributed as $H$ conditioned on $H\le T$. Then $A$ follows the common distribution of node depths of the coalescent point process. We will always assume that $A$ has a density, which we will denote $f$. Recall that $F_T(t):=\bP(A<t)$ and $F(t):= \bP(H<t)$, so that $a=\bP(H>T) = 1-F(T)$ and $F_T(t) = F(t)/F(T)$.

From now on, we will assume that we are given a (reconstructed) phylogenetic tree generated by a coalescent point process, and we will study the change in PD (total length of the tree spanned by the extant species) when $k$ species are uniformly sampled among the $N$ extant species, in particular when this number is random given by a binomial distribution with probability $p$ (the field of bullets model).\\

It has been known since \citet{lam1, lam3} that the reconstructed tree of a wide class of time-continuous binary branching processes (not necessarily Markovian) starting at 0 with one particle and conditioned to survive until time $T$ is a coalescent point process. This statement includes branching processes where the birth rate is possibly time-dependent and the death rate is possibly time-dependent and trait-dependent (if the trait is non-heritable and varies with age and absolute time in the same possibly stochastic manner for all particles).  In the (linear) birth--death process with a constant birth rate $b$ and a constant death rate $d$, this result had been previously known since \citet{ald} in the critical case, and since \citet{ger, ran} in non-critical cases. In these cases, and in the case of time-dependent rates, recall that Equations \eqref{Weq-inhomogeneous}, \eqref{Weqcritical} and \eqref{Weqnotcritical} can be used to plug explicit expressions into the formulae provided hereafter. In the case when the extinction rate additionally depends on age, numerical evaluations can be achieved thanks to Equations \eqref{eqn diff} and \eqref{eqn:h age}.

\section{Expected loss of phylogenetic diversity}

\subsection{Phylogenetic diversity of coalescent point processes}

Recall that the number $N$ of tips in a coalescent point process is a  geometric random variable with success probability $a$
$$
\bP(N=n) = (1-a)^{n-1}a,\qquad n\ge 1.
$$
In addition, the node depths $A_1,\ldots, A_{N-1}$ of the tree are i.i.d. random variable with common distribution function $F_T$, independent of $N$. 

Recall that $S_N$ (respectively $S_n$) denotes the phylogenetic diversity (PD) of the reconstructed tree (respectively of the reconstructed tree conditional on $N=n$), so that the phylogenetic diversity  of a coalescent point process is just 
$$
S_N:= T+\sum_{i=1}^{N-1}A_i.
$$
It will often be more useful to add an exponent $\star$ to denote the PD minus the stem age, provided that the PD is not 0, so that here we have
$$
S_N^\star:= \sum_{i=1}^{N-1}A_i.
$$
Then it is clear that
\begin{equation}
\label{eqn:exp Snstar(1)-1}
\bE(S_n^\star) = (n-1) \bE(A) \quad \mbox{ and }\quad \bE(S_N^\star) = \bE(N-1) \bE(A),
\end{equation}
so that, using $$\bE(A) = \int_0^T \bP(A>t)\, dt,$$ we obtain
\begin{equation}
\label{eqn:exp Snstar(1)-2}
\bE(S_n^\star) = (n-1) \int_0^T (1-F_T(t))\, dt \quad \mbox{ and }\quad \bE(S_N^\star) = a^{-1}(1-a) \int_0^T (1-F_T(t))\, dt.
\end{equation}
Recall that $S_n= T+S_n^\star$, so that $\lim_{n\to\infty}n^{-1} \bE(S_n) = \bE(A)$. Actually, thanks to the strong law of large numbers, we also have the almost sure convergence of $n^{-1} S_n$ to $\bE(A)$.  

Moreover, by the central limit theorem, 
$\sqrt{n}(S_n/n - \bE(A))$ and  $\sqrt{n}(S_n^\star/n - \bE(A))$ both converge in distribution, as $n \rightarrow +\infty$, to a centered Gaussian random variable with the same variance as $A$ (this result is developed further in \citet{cra}).

\subsection{The field of bullets}

\subsubsection{Distribution of the surviving PD when $N$ is random}

Given a coalescent point process with $N$ tips, we remove each of its tips independently with the same probability $1-p$, (the  field of bullets  model \citep{nee, pur, rau}).
We will say that the remaining tips are the `sampled' tips.  The number of sampled tips is denoted by $K$, or by $K_n$ when conditioning on $N=n$. 
We denote by $S_N(p)$ the  PD of the tree spanned by sampled tips, and we set $S_N^\star(p) := S_N(p)-T$ if $K\not=0$ ($0$ otherwise). 

From now on $G$ will denote a  geometric random variable with success probability $p$, that is,
$$
\bP(G=k) = (1-p)^{k-1}p,\qquad k\ge 1.
$$

Let $G'$ denote the random variable equal to ${\rm min} (G, N)$, where $G$ and $N$ are assumed to be independent. Thus $G'$ has a  geometric distribution with success probability $1-(1-p)(1-a)$.  
Let  $A''$ be the maximum of $G'$ independent copies of A, that is
 $$A'':=\max_{i=1,\ldots, G'}A_i$$ where $(A_i)$ are i.i.d. copies of $A$, independent of the geometric random variable $G'$.

The next result expresses the expected values of $S_{N,k}$ and $S_N(p)$ in terms of the expected value of $A''$, and provides explicit formulae for these quantities. Its proof is provided in the Appendix.

\begin{prop}
\label{first}

\begin{equation}
\label{eqn:exp Snstar(p)-1}
\bE(S_{N,k}^\star(p)) = (k-1) \bE(A'')\quad \mbox{ and }\quad \bE(S_N^\star(p)) = \bE((K-1)^+) \bE(A''),
\end{equation}
where $x^+$ denotes the positive part of $x$.  Moreover, the terms in (\ref{eqn:exp Snstar(p)-1}) are given as follows:

\begin{equation}
\label{eqn:exp Snstar(p)-2}
 \bE((K-1)^+) =  \frac{p^2a^{-1}(1-a)}{1-(1-a)(1-p)},
\end{equation}

and

\begin{equation}
\label{eqn:E(A'')}
\bE(A'') = \int_0^T\frac{1-F_T(t)}{1-(1-a)(1-p)F_T(t)}\, dt.
\end{equation}
\end{prop}

Notice that Proposition \ref{first} allows us to recover Eqn. \eqref{eqn:exp Snstar(1)-2} when $p=1$.  
Note also that the first expectation in \eqref{eqn:exp Snstar(p)-1} is conditional on $K=k$ but is not conditional on $N$. 
We will see later that conditioning on $N$ leads to different results with more complicated proofs.

\subsubsection{Expected remaining PD when $N=n$}
\label{expectedPD1}

Here, we assume that $N$ is fixed equal to $n$. We denote by $X_j$ the maximum of $j$ independent copies of $A$. That is, 
$$
X_j:=\max_{i=1,\ldots, j} A_i,
$$
where $(A_i)$ are i.i.d. copies of $A$, so that
we have
\begin{equation}
\label{eqn exp max}
\bE(X_j) =  \int_0^T (1-F_T(t)^j)\ dt.
\end{equation}
The proof of the following result is presented in the Appendix.

\begin{prop}
\label{prop:N=n}
Conditional on $N=n$, we have
$$
\bE(S_n^\star(p))=p^2\sum_{i=2}^n\sum_{j=1}^{i-1}(1-p)^{j-1} \bE(X_j)= p^2\sum_{j=1}^{n-1} (n-j)(1-p)^{j-1}\ \bE(X_j),
$$
where $\bE(X_j)$ is given by \eqref{eqn exp max}, which  can also be expressed as 
$$
\bE(S_n^\star(p))=\frac{p^2}{1-p}\int_0^Tdt\,\big(h_n(1-p) - h_n((1-p)F_T(t))\big),
$$
where 
$$
h_n(x) := \frac{x^{n+1} - nx^2 +(n-1)x}{(1-x)^2} \quad x\not=1.
$$
\end{prop}

Note that by using the second formula in the last statement,  one can easily recover that $\sum_{n\ge 1}a(1-a)^{n-1} \bE(S_n^\star(p)) = \bE(S_N^\star(p))$, where $\bE(S_N^\star(p))$ is given by (\ref{eqn:exp Snstar(p)-1}).

Now let $Y$ be the random variable defined by
$$
Y:= \max_{i=1,\ldots, G}A_i,
$$
where $(A_i)$ represents independent copies of $A$, independent of the geometric random variable $G$ with success probability  $p$, that is,
\begin{equation}
\label{pFTeq}
\bP(Y<t)= \sum_{j\ge 1} p(1-p)^{j-1} \bP(X_j <t) = \sum_{j\ge 1} p(1-p)^{j-1} F_T(t)^j = \frac{pF_T(t)}{1-(1-p)F_T(t)}.
\end{equation}
Also note that $Y$ has the same law as $A''$ when $G'$ is replaced by $G$, so Eqn. (\ref{pFTeq}) also stems from Eqn. \eqref{eqn:law A''} (in the Appendix), when taking a value of $a$ equal to 0, and the following one also stems from Eqn.  \eqref{eqn:E(A'')}:
\begin{equation}
\label{eqn:EY}
\bE(Y)=\int_0^T \bP(Y>t)\, dt = \int_0^T\frac{1-F_T(t)}{1-(1-p)F_T(t)}\, dt.
\end{equation}

The proof of the following result is provided in the Appendix.

\begin{cor}
\label{cor}
We have
$$
\lim_{n\to\infty}\frac{1}{n} \bE(S_n(p)) = p^2\sum_{j=1}^{\infty} (1-p)^{j-1}\ \bE(X_j) = p\bE(Y).
$$
Taking $p=1$, we recover the convergence of $n^{-1} \bE(S_n)$ to $\bE(A)$.
\end{cor}

\subsubsection{Sampling $k$ species out of $n$}
\label{expectedPD2}

In addition to assuming that $N$ is fixed equal to $n$, we now assume that the number $K$ of sampled species also is fixed to $k$, and that these $k$ species are chosen uniformly at random. If the $k$ sampled species are labelled $1,\ldots, k$ from left to right, and if we denote by $C_i$ the coalescence time between sampled species $i$ and sampled species $i+1$, for $i=1,\ldots, k-1$, then it is obvious that $S_{n,k}^\star = \sum_{i=1}^{k-1}C_i$. The next statement yields more information on these new coalescence times. It states that the coalescence times $C_1, \ldots, C_k$ all have the same distribution (moreover, although they are not independent, they are stochastically  `exchangeable',  as is described further in the  Appendix, where the proof of Proposition~\ref{samplingprop} is presented). 

\begin{prop}
\label{samplingprop}
Each coalescence time $C_i$ has the same distribution,  given by 
\begin{eqnarray*}
\bP(C_1<t) &=& k\ \frac{(n-k)!}{n!} \sum_{j=1}^{n-k+1} \frac{(n-j)!}{(n-j-k+1)!} F_T(t)^j\\
	&=& kF_T(t) (1-F_T(t))^{-k}\int_{F_T(t)}^1 y^{n-k}(y-F_T(t))^{k-1}\, dy.
\end{eqnarray*}
\end{prop}

This result allows us to compute the expected value of $S_{n,k}^\star$ as follows.  Recall that $X_j$ denotes the maximum of $j$ independent copies of $A$.
Then the  following statement is an immediate consequence of the last proposition.
\begin{prop}
Conditional on $N=n$ and $K=k$, we have
$$
\bE(S_{n,k}^\star) = k(k-1) \frac{(n-k)!}{n!} \sum_{j=1}^{n-k+1} \frac{(n-j)!}{(n-j-k+1)!} \bE(X_j), 
$$
where $\bE(X_j)$ is given by \eqref{eqn exp max}.
\end{prop}

There is a satisfying connection between this  result and the previous section. Namely, if we compute the expected value of $\bE(S_{n,K}^\star)$ (with respect to the binomial distribution that describes $K$)
then this expected value is  $\sum_{k=1}^n\frac{n!}{k!(n-k)!} p^k (1-p)^{n-k} \bE(S_{n,k}^\star)$;  it is easily verified that this equals $\bE(S_n^\star(p))$, as  given in Proposition \ref{prop:N=n}.

\subsection{Application to predicting biodiversity loss}

Section~\ref{expectedPD1} provides exact expressions for the expected values of $S_n(p)$ (the expected surviving PD) as well as $S_n(1)-S_n(p)$ (the expected loss of PD) in terms of (i) the number $n$ of initial taxa, 
(ii) the depth $T$ of the tree, and (iii) the distribution function $F_T$ for $A$.
To see this, first observe that, by definition:
\begin{equation}
\label{E_star}
\bE(S_n(p)) =\bE(S_n^\star(p)) + T\bP(S_n(p) \not=0)= \bE(S_n^\star(p)) + T(1-(1-p)^n).
\end{equation}

Thus, we have:
$$\bE(S_n(1)-S_n(p)) = \bE(S_n^\star(1)) - \bE(S_n^\star(p)) + T(1-p)^n$$
and the first two terms on the right hand side are given by Proposition~\ref{prop:N=n}.
Similarly, the ratio of expected surviving PD to expected initial PD can be written as 
$(\bE(S_n^\star(1)) +T) /( \bE(S_n^\star(p))+T(1-(1-p)^n)$
and so can also be explicitly determined.  

Similar comments apply for Section~\ref{expectedPD1} if we condition on the number $k$ of taxa that survive the field of bullets extinction event,  since if $k>0$, we have $S_{n}-S_{n,k} = S^\star_n - S^\star_{n,k}$.

\section{Convergence of the ratio of surviving PD to initial PD}

\subsection{Convergence as $n\to\infty$}

For our first result in this section, we allow the full level of generality for the macroevolutionary model that gave rise to the (reconstructed) phylogenetic tree, namely, the speciation rate is possibly time-dependent and the extinction rate is possibly time-dependent and age-dependent (or trait-dependent if the trait is non-heritable and varies in the same, possibly stochastic, manner for all species).

Recall the random variable $Y$ defined by $Y=\max_{i=1,\ldots, G} A_i$, where  $(A_i)$ represents independent copies of $A$, independent of the geometric random variable $G$ with success probability  $p$. Also recall from the beginning of the previous section that $n^{-1} S_n$ converges almost surely to $\bE(A)$ and recall from Corollary \ref{cor} that $n^{-1} \bE(S_n)$ and $n^{-1} \bE(S_n(p))$ converge to $\bE(A)$ and $p\bE(Y)$  respectively.  The proof of the following result is provided in the Appendix. 
\begin{thm} 
\label{thm1}
The following convergence holds almost surely:
$$
\lim_{n\to\infty} n^{-1} S_n(p) = p\bE(Y),
$$
so that 
$$
\pi_T(p) = \frac{p\bE(Y)}{\bE(A)}
$$ 
is not only the limit of $\bE(S_n(p))/\bE(S_n(1))$ but also the almost sure limit of $S_n(p)/S_n(1)$.

For any deterministic sequence $(k_n)$ such that $k_n/n\to p$ as $n\to \infty$, the following convergence holds in probability:
$$
\lim_{n\to\infty} n^{-1} S_{n,k_n} = p\bE(Y),
$$
and $\pi_T(p)$ is both the limit of $\bE(S_{n,k_n})/\bE(S_{n,n})$ and the limit in probability of $S_{n,k_n}/S_{n,n}$.\end{thm}
We point out that $\bE(Y)$ has been computed earlier in the text, in Eqn. (\ref{eqn:EY}):
$$
\bE(Y)= \int_0^T\frac{1-F_T(t)}{1-(1-p)F_T(t)}\, dt.
$$

For the constant--rate birth--death process, in the critical case where the birth and death rates are equal ($b=d$), one can derive an explicit expression for the limiting ratio
$\pi_T(p)$ (which $S_n(p)/S_n(1)$ converges almost surely to),  since Eqns. (\ref{eqn:W}) and (\ref{Weqcritical}) give $F_T(t) = \frac{t(1+bT)}{T(1+bt)}$, from which the following can be derived:

\begin{equation}
\label{explicit}
\pi_T(p) = \frac{p R(x(p))}{R(x(1))},
\end{equation}
where $R(x) = \frac{(1+x) \ln(1+x) -x}{x^2}$ and $x(p) = -1 + p(1+bT)$.
Note that $x(p) > -1$ for all $p>0$  and that $R(x)$ is decreases monotonically with $x \in (-1, +\infty)$  from $R(-1)=1$ to 
an asymptotic value of 0 as $x \rightarrow +\infty$.   In the limit as $bT \rightarrow 0+$, the exact expression for this curve is given by:
\begin{equation}
\label{lim1}
\pi_T(p) = 2p(p\ln(p) + 1-p)/(1-p)^2.
\end{equation}
We will graph this function (and those for other values of $bT$) later in this paper.

\subsection{Convergence as $T\to\infty$}
\label{conv}

Here, we assume that the diversification process is time-homogeneous. Recall that in this case, we denote by $g$ the probability density of species lifetimes, given in particular by \eqref{eqn:lifetime density} in the case when the extinction rate depends on species age. Then the mean number of species begot per mother species is
$$
 m:=b\,\int_0^\infty x\, g(x)\,dx.
$$
We further assume that $m>1$, so the diversification process is supercritical, in the sense that (the survival probability is not zero and that) conditional on the survival event, the number of species increases on average exponentially with time. This is the case for birth-death processes as soon as $r=b-d>0$. Here, we want to get asymptotic results for the ratio of surviving PD  to initial PD as $T$ becomes large. 

Let $Z$ be defined as
$$
Z:=\max_{i=1,\ldots, G} H_i,
$$
where  $(H_i)$ represents independent copies of $H$, independent of the geometric random variable $G$ with success probability $p$. 
The proof of the following theorem is presented in the Appendix. 

\begin{thm} 
\label{thm2}
Conditional on $N_T>0$, we have
$$
\lim_{T\to+\infty} \frac{S_T(p)}{S_T(1)} =  \lim_{T\to+\infty} \frac{\bE(S_T(p))}{\bE(S_T(1))}=\pi_\infty(p),
$$
where the first convergence holds in probability and
$$
\pi_\infty(p)=\frac{p\bE(Z)}{\bE(H)} .
$$
\end{thm}
It is fairly elementary to show in general that
$$
\bE(H) = \int_0^\infty \frac{dt}{W(t)} \quad \mbox{ and }\quad  \bE(Z) = \int_0^\infty \frac{dt}{1-p+pW(t)},
$$
where $W(t) = 1/\bP(H>T)$ ({\em cf.} Eqn.~\ref{Weq}).
As mentioned previously, there is an even  stronger result that the trees obtained from coalescent point processes after Bernoulli sampling with probability $p$ are coalescent point processes with an inverse tail distribution $1-p+pW$ (see Lambert 2011). Recalling that $F$ is the distribution function of $H$, i.e., $1-F= 1/W$, we get
$$
\bE(H) = \int_0^\infty (1-F(t))\,dt \quad \mbox{ and }\quad  \bE(Z) = \int_0^\infty \frac{1-F(t)}{1-(1-p)F(t)}\, dt,
$$
which is the form chosen to display Theorem \ref{thm2} in the introduction (Eqn. \eqref{eqn:main2}).

Now let us do the calculations in the case of birth--death processes. First,
$$
\bE(H) = \int_0^\infty \frac{dt}{1+\frac{b}{r}(e^{rt}-1)} = \left[d^{-1}\ln\left(b-de^{-rt}\right)\right]_0^\infty = \frac{1}{d} \ln\left(\frac{b}{r}\right),
$$
which simply equals $b^{-1}$ in the Yule case when $d=0$ (since then $H$ is exponentially distributed with parameter $b$). Second, $1-p+pW(t) = 1+\frac{bp}{r}(e^{rt}-1)$, so we get the same expression as for $W$ after replacing $b$ with $bp$ but keeping $r$ unchanged.
As a result,
$$
\bE(Z)  =\frac{1}{bp-r} \ln\left(\frac{bp}{r}\right) ,
$$
which simply equals $r^{-1}$ when $p= r/b$. This can be recorded in the following corollary.
\begin{cor}
\label{coro}
 In the case of birth--death processes with speciation rate $b$, extinction rate $d$, 
and diversification rate $r=b-d>0$, 
$$
\pi_\infty(p) =
\begin{cases}
\frac{dp}{bp-r}\frac{\ln(bp/r)}{\ln (b/r)}, & \mbox{ if }  b>r\not= bp;\\
 -\frac{p\ln (p)}{1-p}, &  \mbox{ if }  b=r> bp; \\
-\frac{1-p}{\ln (p)}, & \mbox{ if }  b>r= bp.
\end{cases}
$$
\end{cor}

\subsection{Additional remarks}

\begin{itemize}
\item  {\em Proportion of the PD in the original tree or in another tree?}

We have established the convergence of the random ratios $S_n(p)/S_n(1)$ and $S_T(p)/S_T(1)$ to the constants $\pi_T(p)$ and $\pi_\infty(p)$, respectively.
Actually, we do not directly prove the convergence of these quantities  seen as the ratios of new-to-old PD of the \emph{same} original (random) tree. Strictly speaking, we prove that each PD (new vs old) \emph{separately}, converges, after normalization by the \emph{same} quantity. 
For example, we prove the convergences of $S_n(p)/n$ and of $S_n(1)/n$. 
Because these two convergences are actually laws of large numbers, both limits are deterministic, and the convergence in probability of each normalized PD ($S_n(p)/n$ and $S_n(1)/n$) implies the convergence in probability of their ratio to the ratio of their limits. And since each PD is normalized by the same quantity, the ratio of normalized PDs is also 
the ratio of the new-to-old PD of the same (random) tree.
The study of fluctuations around the limiting value (see next point) is more problematic, because we use the central limit theorem and have to deal with covariances between new and old PD.

\item {\em Distribution of $S_T(p)/S_T(1)$ about its limiting value.}

  Although Theorem~\ref{thm1} ensures the almost sure convergence of $S_n(p)/S_n(1)$ to its limit, for applications it is useful to also have on hand the distribution of  $S_n(p)/S_n(1)$ about its limit,  for finite values of $n$.  It can be shown (see Appendix, Subsection \ref{subsec:clt}) that this distribution is asymptotically normally distributed with a standard deviation that can be explicitly computed, and which decays towards zero at the rate $1/\sqrt{n}$. More specifically, let $G$ be a geometric random variable  with success probability  $p$, let $(A_i)$ be independent copies of the typical node depth $A$ and set
$$
Y:= \max_{i=1,\ldots, G}A_i, \quad \mbox{ and }\quad
Y':= \sum_{i=1}^GA_i,
$$
where both definitions use the same $G$ and the same $(A_i)$. In particular, $Y\le Y'$. Then  
$$
\lim_{n\to\infty}\sqrt{n}\left(\frac{S_{n}(p)}{S_{n}(1)} -  \pi_T(p)\right) = p^{-1/2}V,
$$
where $V$ is a centered, Gaussian random variable with variance $\sigma^2$ equal to
\begin{multline}
\label{sigma}
\sigma^2 = \left(\frac{\bE(Y)}{\bE(Y')}\right)^2 \text{Var} \left( \frac{Y}{\bE(Y)} - \frac{Y'}{\bE(Y')} \right) \\
=  \left(\frac{\bE(Y)}{\bE(Y')}\right)^2 \,\left(
\frac{\bE(Y^2)}{\bE(Y)^2} + \frac{\bE(Y'^2)}{\bE(Y')^2} -2 \frac{\bE(YY')}{\bE(Y)\bE(Y')}
\right) .
\end{multline}
This gives us a quantification of the error made by approximating the ratio of new-to-old PD by the asymptotic ratio $\pi_T$. Roughly speaking, this error is of the order $\sigma/\sqrt{pn}$. Moreover, $\sigma$ can be explicitly computed (see in particular Lemma \ref{last one} in the Appendix).

\end{itemize}

\section{Properties of the phylogenetic diversity ratios ($\pi(p)$)}

In the previous section, we have been able to provide a closed-form formula for the limiting proportion of PD that would remain after mass extinction events at the present for trees generated under the wide class of models described by \citet{lam1, lam3}. Here, `limiting' refers to the case when the number of species $n$ becomes large for a fixed value of $T$, or when $T$ becomes large for a supercritical process conditioned on non-extinction. In both cases, when each species survives independently with probability $p$, this limiting proportion can be expressed under the general form
\begin{equation}
\label{pdratio}
\pi(p)= \frac{p\int_I \frac{1-G(t)}{1-(1-p)G(t)}\, dt}{ \int_I (1-G(t))\, dt},
\end{equation}
where the notation ($I, G$ and $\pi$) is as follows.  In the first case, where $\pi=\pi_T$, $I$ denotes the interval $[0,T]$ and $G=F_T$, i.e., $G(t) = F(t)/F(T)$. In the second case, where $\pi=\pi_\infty$, $I$ denotes the interval $[0,\infty)$ and $G=F$. We will now refer to  $\pi(p)$ as the {\em phylogenetic diversity ratio}. In this section we describe some of the properties of the mapping $p \mapsto \pi(p)$.

Firstly, notice that from Eqns~(\ref{eqn:main1}) and (\ref{eqn:main2}), this mapping is fully determined by just $p$ and the distribution $F$ associated with the underlying coalescent point process for the model.

Secondly,  it is clear that $\pi(0)=0$, $\pi(1)=1$; moreover, $\pi(p)\geq p$ for all $p \in [0,1]$ (this last fact can be shown directly from
Eqn. (\ref{pdratio}) or from results that follow).  Thus, the proportion of lost PD under these models will always  (in the limit) be less than the proportion of present-day taxa that disappear.

 Furthermore, $p\mapsto\pi(p)$ is a strictly increasing function of $p$.  This last fact can be verified by various arguments, including a direct appeal to Eqns~(\ref{eqn:main1}) and (\ref{eqn:main2}). Using the Leibniz rule to differentiate (wrt $p$) inside the integral and re-arranging, we obtain 
\begin{equation}
\label{firstderiv}
\frac{d}{dp}\pi(p)= \frac{\int_I\left[\frac{ 1-G(t)}{ 1-(1-p)G(t)}\right]^2 dt}{\int_I (1-G(t))dt}\qquad p\in(0,1).
\end{equation}
Notice that in the first case, when $T<\infty$, as $p \rightarrow 0+$, this last equation gives
\begin{equation}
\label{pto0}
\lim_{p \rightarrow 0+} \frac{d}{dp}\pi_T(p)
 = \frac{T}{\int_0^T (1-G(t))dt},
\end{equation}
which is a sublinear function of  $T$, and which tends to $+\infty$ as $T \rightarrow \infty$. In the second case, it is obvious that $\lim_{p \rightarrow 0+} \frac{d}{dp}\pi_\infty(p) = +\infty$.

The limiting value $\lim_{p \rightarrow 1-} \frac{d}{dp}\pi(p)$ is more interesting. Formally,
\begin{equation}
\label{pto1}
\lim_{p \rightarrow 1-} \frac{d}{dp}\pi(p)= \frac{\int_I (1-G(t))^2 dt}{\int_I (1-G(t))dt} = 1- \frac{\int_I G(t)(1-G(t)) dt}{\int_I (1-G(t))dt} .
\end{equation}
We can also provide an explicit interpretation of this value by considering the field of bullets operating on any \emph{given} binary tree  $\cT$ with positive branch lengths, as we now explain.  Let  $S(\cT,p)$ denote the PD remaining after the passing of the field-of-bullets model (with leaf survival probability $p$) on this tree. 
As in  \citet{fal}, we can write:
 $$\bE_{\cT}(S(\cT, p))  = S(\cT, 1) - \sum_{e}l_e (1-p)^{n(e)},$$ where $\bE_{\cT}$ denotes expectation with respect to the passage of the field of bullets model, conditional on the given tree $\cT$, the summation is over all the edges of $\cT$, $l(e)$ is the length of edge $e$ and
$n(e)$ is the number of leaves of  $\cT$  that are descendants of $e$.  
It follows that
\begin{equation}
\label{firstd}
\frac{d}{dp}\bE_{\cT}(S(\cT, p))=  \sum_{e} l(e)n(e) (1-p)^{n(e)-1} .
\end{equation}
Now, we let $\bE$ denote the law of $\cT$ and we assume it is that of a coalescent point process, conditioned or not upon its number of tips. Letting  $S(p)$ denote the corresponding PD, an application of the Dominated Convergence Theorem yields
$$ \frac{d}{dp}\bE(S (p)) = \frac{d}{dp}\bE( \bE_{\cT}(S(\cT, p))) = \bE\left(\frac{d}{dp} \bE_{\cT}(S(\cT, p))\right).$$
Notice that as $p \rightarrow 1-$ the right-hand side of (\ref{firstd}) converges to $\sum_{e: n(e)=1} l(e)$, which is the sum of the lengths of the pendant edges, and so 
\begin{equation}
\label{dag1}
 \lim_{p \rightarrow 1-} \frac{d}{dp}\bE(S (p)) =  \bE\left(\sum_{e: n(e)=1} l(e)\right).
 \end{equation}

It can now be seen that $\lim_{p \rightarrow 1-} \frac{d}{dp}\pi(p)$ is the (limiting) ratio of the expected PD that is spanned by the pendant branches, divided by the  expected total PD in the tree. For example, for  the constant--rate pure--birth model, the ratio of the expected sum of the pendant branch lengths to the expected sum of the interior (non-pendant) branch lengths converges to $1$ as $T \rightarrow +\infty$ \citep{moo} 
 and so $\lim_{p \rightarrow 1-} \frac{d}{dp}\pi_\infty(p) = \frac{1}{2}$.

Another generic property of the phylogenetic diversity ratio is that it is strictly concave:
\begin{equation}
\label{concave}
\frac{d^2}{dp^2}\pi(p)<0, \qquad p\in(0,1).
\end{equation}
This is evident from the following identity, which is obtained by a further differentiation (with respect to $p$) of Eqn. (\ref{firstderiv}):
\begin{equation}
\label{secondderiv}
\frac{d^2}{dp^2}\pi(p)= -2 \frac{\int_I  \frac{ (1-G(t))^2G(t)}{( 1-(1-p)G(t))^3} dt}{\int_I (1-G(t))dt}
\end{equation}
The concave relationship (Eqn.~(\ref{concave})) is also a consequence of the concavity of expected PD on a fixed binary tree $\cT$ with positive branch lengths.  More precisely, by further differentiation (with respect to $p$) of Eqn. (\ref{firstd}), we obtain
\begin{equation}
\label{d2eq}
\frac{d^2}{dp^2}\bE_{\cT}(S(\cT, p))= - \sum_{e: n(e)>1} l(e)n(e)(n(e)-1) (1-p)^{n(e)-2} \leq 0.
\end{equation}
Notice that the  inequality in Eqn. (\ref{d2eq}) is strict unless all of the interior edges of $\cT$ have zero length.  Consequently, taking expectation with 
respect to the tree $\cT$ and its branch lengths, one could also recover Eqn. (\ref{concave}) from Eqn. (\ref{d2eq}).

\subsection*{Extreme cases}

\begin{figure}
\centering
\mbox{\subfigure{\includegraphics[width=3in]{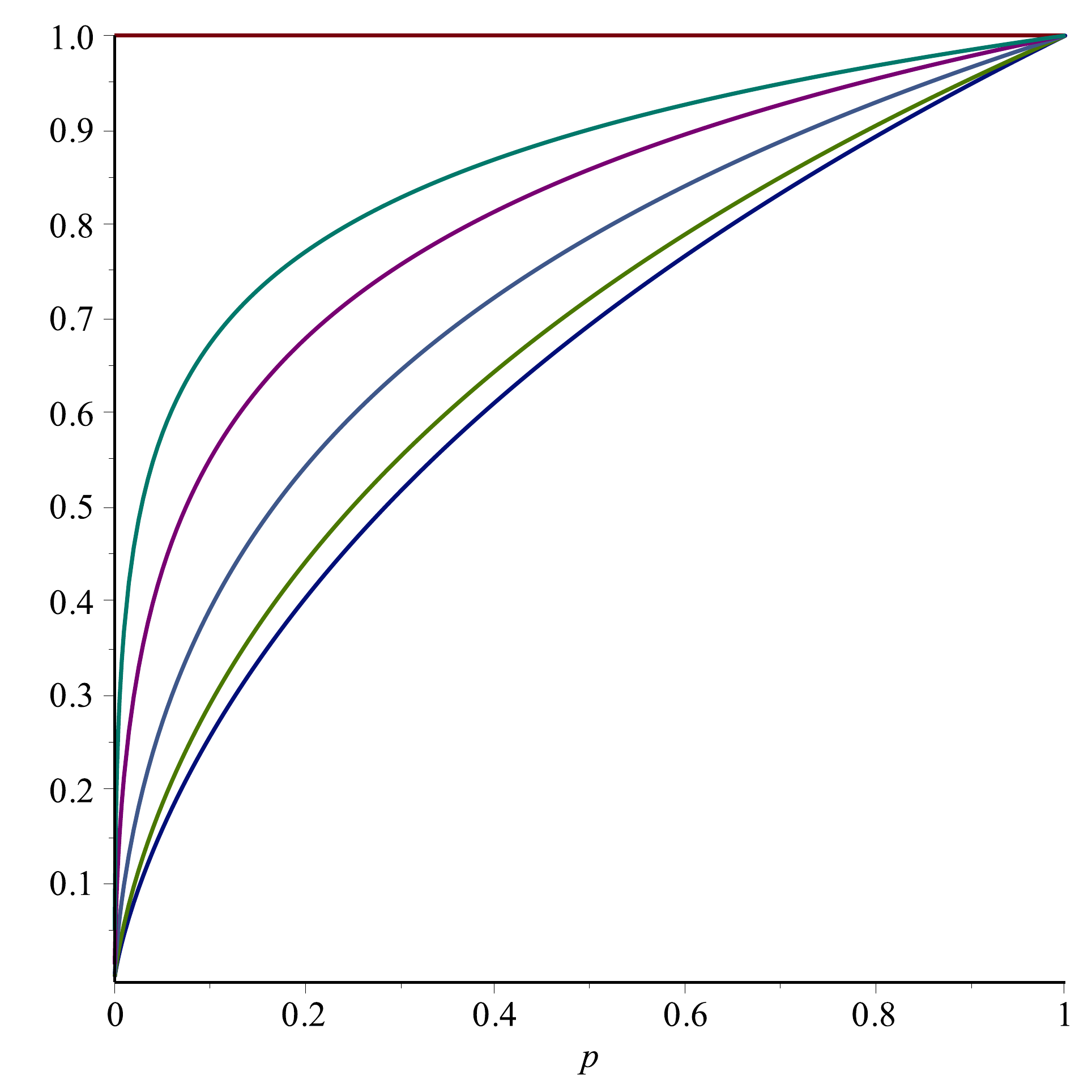}}\quad
\subfigure{\includegraphics[width=3in]{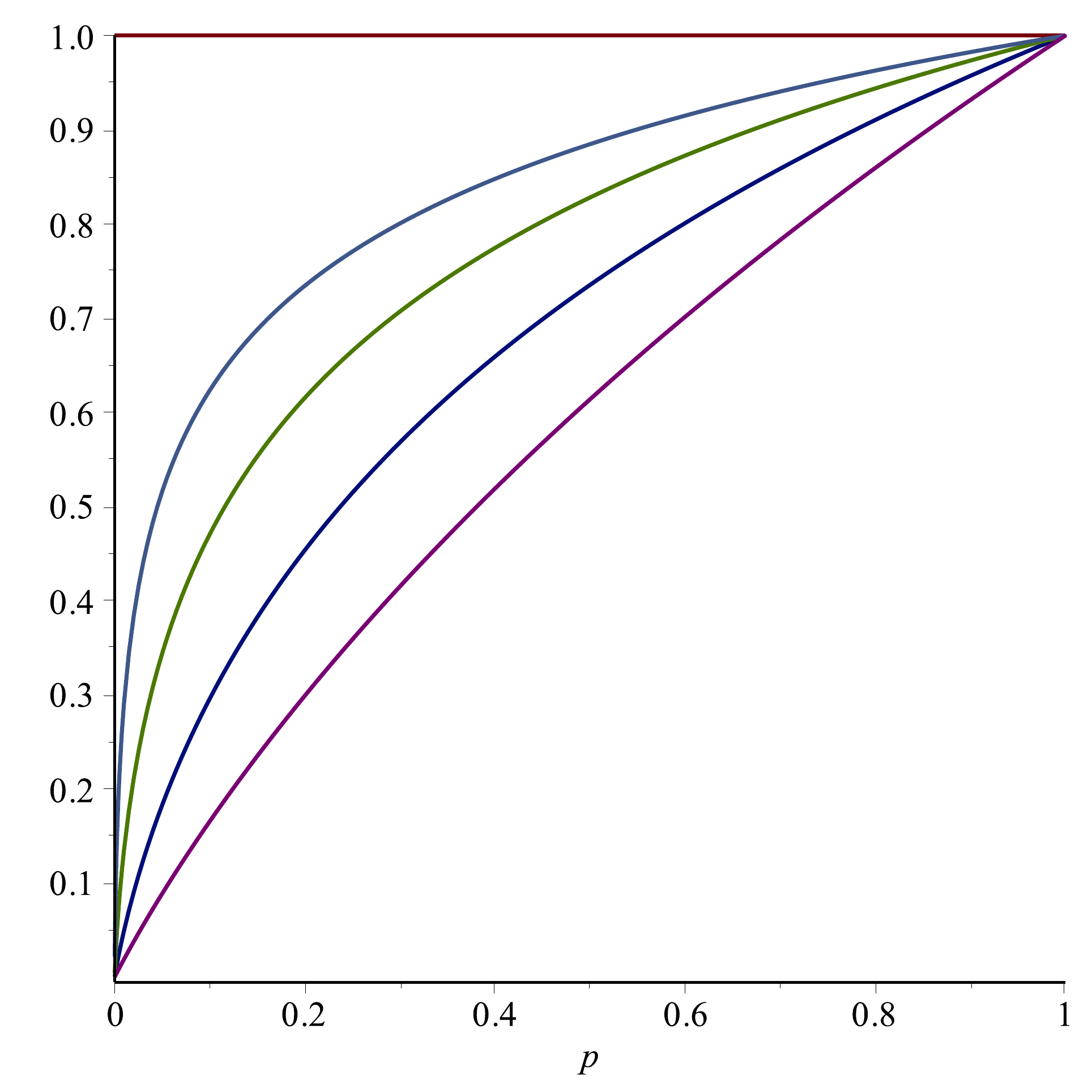} }}
\caption{Left: The slow progression of the curve $p \mapsto \pi_\infty(p)$ towards the step function $U$ (top line) for a constant--rate birth--death process for $d/b = 0$ (the lowest curve)and $d/b= 0.5, 0.9, 0.99, 0.999.$ Right: For a critical constant--rate birth--death process, the graph of $p \mapsto \pi_T(p)$ also shows a slow progression towards $U$ as $bT \rightarrow \infty$. Here, $bT \rightarrow 0+$ is the lowest curve, with $bT=10, 100, 1000$ for the curves of increasing height.} \label{comparison_fig}
\end{figure}

Although the curve $p \mapsto \pi_T(p)$ lies above the straight line $p \mapsto p$, there are birth-death processes for which the PD ratio curve comes arbitrarily close to this straight line. 
Consider first the constant--rate pure-birth model ($b(t)=b, d=0$) for which we have:
$$\pi_T(p) =p \int_0^T \frac{e^{-bt}-e^{-bT}}{(1-p)e^{-bt} -e^{-bT}+p}dt.$$
This function converges to $-p\ln(p)/(1-p)$ as $T$ tends to infinity, shown as the lowest curve in the left-hand graph of Fig.~\ref{comparison_fig} (and in agreement with Corollary~\ref{coro}).
Now compare this with a pure--birth model where the speciation rate is  (exponentially) higher in early times; in this case  we obtain a curve that can be made as close to the line $p \mapsto p$ as we wish.
More precisely, if we set
$$b(t)=\gamma e^{-\gamma t} \mbox{ and }  d=0,$$
then Eqn. (\ref{Weq-inhomogeneous}) shows that in the limit as $\gamma \rightarrow \infty$ we have:
\begin{equation}
\label{lim2}
\pi_T(p) = p.
\end{equation}  
That is, the phylogenetic diversity ratio is (asymptotically) proportional to the expected proportion of surviving species. This makes perfect sense, since as $\gamma$ becomes large, all speciation events occur right at the start of the tree, thereby generating a star-like phylogeny.

A further limiting result concerns the behavior of $\pi(p)$ as we approach a critical  process where the extinction rate equals the speciation rate.
Corollary~\ref{coro} implies that for a supercritical constant--rate birth--death process ($r>0$), we have:
\begin{equation}
\label{limit1}
\lim_{r \rightarrow 0+} \pi_{\infty}(p) = U(p).
\end{equation}
where $U$ is the step function defined by:
\begin{equation}
\label{step}
U(p) = 
\begin{cases}
0, & \mbox{ for }  p=0;\\
1, &  \mbox{ for }  0<p\leq 1.
\end{cases}
\end{equation}

The left-hand graph in Fig.~\ref{comparison_fig} illustrates this rather slow convergence towards $U(p)$ as $r$ decreases towards $0$. 

Also, by Eqn.~(\ref{explicit}), when $r=0$ (i.e. a critical constant--rate birth--death process), we also 
have:
\begin{equation}
\label{limit2}
\lim_{T \rightarrow +\infty} \pi_{T}(p) = U(p).
\end{equation}
Note that in both these two limits, the expected node depth $H$ in the coalescent point process is diverging to $+\infty$ (as $r \rightarrow 0$ in (\ref{limit1}) and, by  Eqn.~(\ref{Weqcritical}), as $T \rightarrow +\infty$ in Eqn. (\ref{limit2})).  Moreover, when $r=0$, the tree is guaranteed to become extinct as $T \rightarrow \infty$; therefore,  the limit in Eqn. (\ref{limit2}), while formally correct, is not particularly meaningful as it involves conditioning on an event that has a limiting probability of zero.  The graph of the curve $p \mapsto \pi_T(p)$ for this critical constant--rate birth--death process, is shown in the right-hand graph of Fig.~\ref{comparison_fig} for $bT \rightarrow 0+$ (the lowest curve, described exactly by Eqn. (\ref{lim1})), and $bT=10, 100, 1000$. 

As a final remark, notice that  $\frac{d}{dp}\pi(p)$ is strictly greater than 1 as $p \rightarrow 0+$ (by Eqn. (\ref{pto0})) and is strictly less than 1 as $p \rightarrow 1-$ (by Eqn. (\ref{pto1})), and so, since $\frac{d}{dp}\pi(p)$   strictly decreases between $0$ and $1$ (by the concavity relationship in Eqn. (\ref{concave})), there is a unique value $p=p^*$ 
for which
\begin{equation}
\label{tipping}
\frac{d}{dp} \pi(p^*) = 1.
\end{equation}
Moreover, for all  $p<p^*$,  $\pi(p)$ has a super--linear dependence on $p$, while for all $p>p^*$, the dependence is sub-linear.
Solving Eqn. (\ref{tipping}) for $p^*$ provides the precise transition point between these two regimes. For example, for the constant--rate pure--birth model (the case where $b=r>bp$ in Corollary~\ref{coro}), we have $\pi_\infty(p) = \frac{-p\ln(p)}{1-p}$;  routine calculus shows that $p^*$ is the solution to the equation
$-\ln(p) = (2-p)(1-p)$ for $0<p <1$, which gives us $p^* \approx  0.316$.

An interesting question is whether every supercritical birth--death process, with a constant speciation rate and an extinction rate that is dependent only on age, leads to a ratio $\pi_\infty(p)$ that always lies above the curve $-p\ln(p)/(1-p)$ of a pure-birth process.

\section{Concluding comments}

Measuring biological diversity in terms of evolutionary heritage is a type of `last hope' one might seek to cling to in the face of the current biodiversity crisis.  In other words, despite many extinction events, we might still keep most of the biodiversity, because the loss of a species that is closely related to other species that survive  results in a relatively small decline in phylogenetic diversity.    Some minor mathematical support for this view is provided by the inequality  $\pi(p) \geq p$ (the loss of relative PD under the field of bullets model is never more than the loss of relative species numbers); however, we have also shown that for certain (time-inhomogeneous) diversification models, the function $\pi(p)$ can be as close to $p$ as we wish.  

At the other extreme, we have seen that there are diversification models for which $\pi$ approaches the step function $U$ (with $\pi(p)=1$ for $p>0$).  In a well-cited paper, Nee and May (1997) stated that ``80\% of the underlying tree of life can survive even when approximately 95\% of species are lost.''  However, their analysis involved trees produced by Kingman's coalescent model from population genetics, which results in tree shapes
with extremely short pendant edges and a few very long deep interior edges, as noted by \citet{moo}.  By contrast, the shape of most real evolutionary trees tend to be better described by models that are closer to the shape of pure--birth trees than Kingman coalescent trees  (see e.g. \citet{hey}, \citet{mcP},  \citet{mor}). In  other words the data tend to fit models where $b\gg d$ better than $b=d$.
 In this case, $\pi(p)$ lies much closer to the curve $-p\ln(p)/(1-p)$ of a pure--birth process, for which the loss of $95\%$ of species would lead to the {\em loss} of more than $84\%$ of the PD \citep{moo}.  

Given that extinction plays such a major role in evolution \citep{erw}, it may seem surprising that reconstructed trees tend to fit models with high values of $b$ relative to $d$.
However, ascertainment bias may provide an explanation for this. 
For example, as $d$ gets close to $b$, we get a tree which is extremely 
unlikely to have survived (and additionally has node depths with 
infinite expectations), so this extreme world where $\pi= U$ is totally 
unlikely. In the real world, phylogenetic trees are much more likely to 
have a small $d$ and a high $b$, just because they have survived, so is it more 
likely to get a $\pi$ which is closer to the lower curve $-p\ln(p)/(1-p)$. In other words, clades that had $b \approx d$ would have been much less likely to have
left any surviving taxa today than clades with $b\gg d$.  

The fact that $\pi$ is always concave as a function of $p$ (including if we condition on the underlying tree) means that if the present mass extinctions are 
more or less painless (slow and sublinear), they will  go faster than expected below $p^*$ when the dependence becomes super-linear.
Moreover, the field of  bullets model represents a conservative estimate:  in reality, the extinction of a 
contemporary species will be likely to be correlated with the extinction of closely related species (due to shared traits \citep{fal2} or niche proximity), which 
would yield an even worse portrait than the one we depict.

Notice that we have two extinctions in our analysis -- a rapid (mass) extinction at the present (modeled by the field of bullets model)  and the extinctions that are part of the on-going and slower rate diversification process that generates the  phylogenetic tree.  There is a good reason to treat these two processes separately -- the extinctions at the present are considered to be over a time-scale that is effectively instantaneous on an evolutionary time-scale (e.g. 100 years)  and too short for new speciation events, or for existing branch lengths to change significantly.

Danish physicist Neils Bohr (1885--1962)  is credited with the quip ``Prediction is very difficult, especially about the future."    In the case of predicting potential biodiversity loss in the near future,  not  only is a rapid extinction event likely to be a  highly random process, but the resulting loss of biodiversity depends also on the properties of the underlying evolutionary tree, only some of which are known with any precision.  We have shown that  under a general class of diversification models and a simple model of mass extinction,  the proportion of lost PD can be estimated from two simple quantities:  the expected proportion of taxa that survive ($p$) and the distribution of coalescence times ($F_T$).    The latter function may be estimated from the shapes of 
reconstructed trees, reflected in the way in which branch lengths are distributed.  For certain data-sets, early radiations followed by long periods of stasis lead to quite different 
shaped trees from ones in which recent speciation rates are higher ({\em cf. } \citet{mor}, \citet{rab}).   It would be of interest to estimate $F_T$  for a variety of real data sets, and to determine the impact of tree shape on expected biodiversity loss, for various values of $p$.

One feature of our approach is that it allows for general properties (and upper and lower bounds)  of biodiversity loss to be determined, as well as the estimation, for any particular model, of how much diversity is likely to disappear as a function of $p$.  In that sense, the explicit expressions for $\pi(p)$ may be viewed as biodiversity analogues of some early formulae in population genetics concerning allele frequencies\footnote{Curiously, the same function $-p\ln(p)/(1-p)$ that appears in Corollary~\ref{cor} also plays a role in population genetics for the estimation of the mean time till the loss of a deleterious allele where the initial proportion of the allele is $p$.}.  The stronger of our two convergence results (Theorem~\ref{thm1}) is the case where the time-scale $T$ is fixed and $n$ grows. In  this case, we require the least restriction on the diversification process (the speciation rate can depend on time, and the extinction rate can depend on both time and lineage age), and in this case we have almost sure convergence of the proportion of surviving PD  to its expected value rather than just convergence in probability.   Our second convergence result (Theorem~\ref{thm2}) holds for super--critical processes with a constant speciation rate (but an extinction rate that may depend on time and lineage age) and holds in the limit as $T$ becomes large.  

Finally, we note that, despite Bohr's quip above, the almost sure convergence of phylogenetic ratio to $\pi_T(p)$ provides a considerable bonus over merely computing expected values -- it  shows that PD loss becomes more predictable than we might imagine, at least under the models we have investigated here.

\paragraph{Acknowledgments.}

We thank Arne Mooers for some helpful suggestions on an earlier version of this manuscript. AL was financially supported by grant MANEGE `Mod\`eles Al\'eatoires en \'Ecologie, G\'en\'etique et \'Evolution' 09-BLAN-0215 of ANR (French national research agency). AL also thanks the {\em Center for Interdisciplinary Research in Biology} (Collège de France) for funding.
MS thanks the {\em Allan Wilson Centre for Molecular Ecology and Evolution} for funding.

\bibliographystyle{model2-names}
\bibliography{JTB_Amaury_Steel}

\begin{thebibliography}{20}
\expandafter\ifx\csname natexlab\endcsname\relax\def\natexlab#1{#1}\fi
\providecommand{\url}[1]{\texttt{#1}}
\providecommand{\href}[2]{#2}
\providecommand{\path}[1]{#1}
\providecommand{\DOIprefix}{doi:}
\providecommand{\ArXivprefix}{arXiv:}
\providecommand{\URLprefix}{URL: }
\providecommand{\Pubmedprefix}{pmid:}
\providecommand{\doi}[1]{\href{http://dx.doi.org/#1}{\path{#1}}}
\providecommand{\Pubmed}[1]{\href{pmid:#1}{\path{#1}}}
\providecommand{\bibinfo}[2]{#2}
\ifx\xfnm\relax \def\xfnm[#1]{\unskip,\space#1}\fi
\bibitem[{Aldous et~al.(2011)Aldous, Krikun and Popovic}]{ald3}
\bibinfo{author}{Aldous, D.}, \bibinfo{author}{Krikun, M.},
  \bibinfo{author}{Popovic, L.}, \bibinfo{year}{2011}.
\newblock \bibinfo{title}{Five statistical questions about the tree of life}.
\newblock \bibinfo{journal}{Syst. Biol.} \bibinfo{volume}{60},
  \bibinfo{pages}{318--328}.
\bibitem[{Aldous and Popovic(2005)}]{ald}
\bibinfo{author}{Aldous, D.}, \bibinfo{author}{Popovic, L.},
  \bibinfo{year}{2005}.
\newblock \bibinfo{title}{A critical branching process model for biodiversity}.
\newblock \bibinfo{journal}{Adv. Appl. Probab.} \bibinfo{volume}{37},
  \bibinfo{pages}{1094--1115}.
\bibitem[{Crawford and Suchard(2012)}]{cra}
\bibinfo{author}{Crawford, F.}, \bibinfo{author}{Suchard, M.A.},
  \bibinfo{year}{2012}.
\newblock \bibinfo{title}{Diversity, disparity, and evolutionary rate
  estimation for unresolved {Y}ule trees}.
\newblock \bibinfo{journal}{Syst. Biol.} \bibinfo{volume}{62},
  \bibinfo{pages}{439--455}.
\bibitem[{Erwin(2008)}]{erw}
\bibinfo{author}{Erwin, D.}, \bibinfo{year}{2008}.
\newblock \bibinfo{title}{Extinction as the loss of evolutionary history}.
\newblock \bibinfo{journal}{P. Natl. Acad. Sci. USA} \bibinfo{volume}{105},
  \bibinfo{pages}{11520--11527}.
\bibitem[{Faith(1992)}]{fai}
\bibinfo{author}{Faith, D.}, \bibinfo{year}{1992}.
\newblock \bibinfo{title}{Conservation evaluation and phylogenetic diversity}.
\newblock \bibinfo{journal}{Biol. Conserv.} \bibinfo{volume}{61},
  \bibinfo{pages}{1--10}.
\bibitem[{Faller et~al.(2008)Faller, Pardi and Steel}]{fal}
\bibinfo{author}{Faller, B.}, \bibinfo{author}{Pardi, F.},
  \bibinfo{author}{Steel, M.}, \bibinfo{year}{2008}.
\newblock \bibinfo{title}{Distribution of phylogenetic diversity under random
  extinction}.
\newblock \bibinfo{journal}{J. Theor. Biol.} \bibinfo{volume}{251},
  \bibinfo{pages}{286--296}.
\bibitem[{Faller and Steel(2012)}]{fal2}
\bibinfo{author}{Faller, B.}, \bibinfo{author}{Steel, M.},
  \bibinfo{year}{2012}.
\newblock \bibinfo{title}{Trait-dependent extinction leads to greater expected
  biodiversity loss}.
\newblock \bibinfo{journal}{SIAM J. Discrete Math.} \bibinfo{volume}{26},
  \bibinfo{pages}{472--482}.
\bibitem[{Geiger and Kersting(1997)}]{gei}
\bibinfo{author}{Geiger, J.}, \bibinfo{author}{Kersting, G.},
  \bibinfo{year}{1997}.
\newblock \bibinfo{title}{Depth-first search of random trees, and poisson point
  processes}, in: \bibinfo{booktitle}{Classical and Modern Branching Processes,
  IMA Math. Appl. 84.}, \bibinfo{publisher}{Springer, New York},
  \bibinfo{address}{Minneapolis, MN, USA}. pp. \bibinfo{pages}{111--126}.
\bibitem[{Gernhard(2008)}]{ger}
\bibinfo{author}{Gernhard, T.}, \bibinfo{year}{2008}.
\newblock \bibinfo{title}{The conditioned reconstructed process}.
\newblock \bibinfo{journal}{J. Theor. Biol.} \bibinfo{volume}{253},
  \bibinfo{pages}{769--778}.
\bibitem[{Hey(1992)}]{hey}
\bibinfo{author}{Hey, J.}, \bibinfo{year}{1992}.
\newblock \bibinfo{title}{Using phylogenetic trees to study speciation and
  extinction}.
\newblock \bibinfo{journal}{Evolution} \bibinfo{volume}{46},
  \bibinfo{pages}{627--640}.
\bibitem[{Lambert(2010)}]{lam1}
\bibinfo{author}{Lambert, A.}, \bibinfo{year}{2010}.
\newblock \bibinfo{title}{The contour of splitting trees is a l{\'e}vy
  process}.
\newblock \bibinfo{journal}{Ann. Probab.} \bibinfo{volume}{38},
  \bibinfo{pages}{348--395}.
\bibitem[{Lambert and Stadler(2013)}]{lam3}
\bibinfo{author}{Lambert, A.}, \bibinfo{author}{Stadler, T.},
  \bibinfo{year}{2013}.
\newblock \bibinfo{title}{Macro-evolutionary models and coalescent point
  processes: the shape and probability of reconstructed phylogenies}.
\newblock \bibinfo{journal}{(Submitted)} .
\bibitem[{McPeek(2008)}]{mcP}
\bibinfo{author}{McPeek, M.}, \bibinfo{year}{2008}.
\newblock \bibinfo{title}{The ecological dynamics of clade diversification and
  community assembly}.
\newblock \bibinfo{journal}{Am. Nat.} \bibinfo{volume}{172},
  \bibinfo{pages}{E270--E284}.
\bibitem[{Mooers et~al.(2012)Mooers, Gascuel, Stadler, Li and Steel}]{moo}
\bibinfo{author}{Mooers, A.}, \bibinfo{author}{Gascuel, O.},
  \bibinfo{author}{Stadler, T.}, \bibinfo{author}{Li, H.},
  \bibinfo{author}{Steel, M.}, \bibinfo{year}{2012}.
\newblock \bibinfo{title}{Branch lengths on {Y}ule trees and the expected loss
  of phylogenetic diversity}.
\newblock \bibinfo{journal}{Syst. Biol.} \bibinfo{volume}{61},
  \bibinfo{pages}{195--203}.
\bibitem[{Morlon et~al.(2010)Morlon, Potts and Plotkin}]{mor}
\bibinfo{author}{Morlon, H.}, \bibinfo{author}{Potts, M.},
  \bibinfo{author}{Plotkin, J.}, \bibinfo{year}{2010}.
\newblock \bibinfo{title}{Inferring the dynamics of diversification: A
  coalescent approach}.
\newblock \bibinfo{journal}{PLoS Biol.} \bibinfo{volume}{8},
  \bibinfo{pages}{e1000493}.
\newblock \bibinfo{note}{Doi:10.1371/journal.pbio.1000493}.
\bibitem[{Nee and May(1997)}]{nee}
\bibinfo{author}{Nee, S.}, \bibinfo{author}{May, R.}, \bibinfo{year}{1997}.
\newblock \bibinfo{title}{Extinction and the loss of evolutionary history}.
\newblock \bibinfo{journal}{Science} \bibinfo{volume}{278},
  \bibinfo{pages}{692--694}.
\bibitem[{Purvis et~al.(2000)Purvis, Agapow, Gittleman and G.M.}]{pur}
\bibinfo{author}{Purvis, A.}, \bibinfo{author}{Agapow, P.},
  \bibinfo{author}{Gittleman, J.}, \bibinfo{author}{G.M., M.},
  \bibinfo{year}{2000}.
\newblock \bibinfo{title}{Nonrandom extinction and the loss of evolutionary
  history}.
\newblock \bibinfo{journal}{Science} \bibinfo{volume}{288},
  \bibinfo{pages}{328--330}.
\bibitem[{Rabosky and Lovette(2008)}]{rab}
\bibinfo{author}{Rabosky, D.}, \bibinfo{author}{Lovette, I.},
  \bibinfo{year}{2008}.
\newblock \bibinfo{title}{Density-dependent diversification in north american
  wood warblers}.
\newblock \bibinfo{journal}{P. Roy. Soc. B Biol.} \bibinfo{volume}{275},
  \bibinfo{pages}{2363--2371}.
\bibitem[{Rannala(1997)}]{ran}
\bibinfo{author}{Rannala, B.}, \bibinfo{year}{1997}.
\newblock \bibinfo{title}{Gene genealogy in a population of variable size}.
\newblock \bibinfo{journal}{Heredity} \bibinfo{volume}{78},
  \bibinfo{pages}{417--423}.
\bibitem[{Raup(1992)}]{rau}
\bibinfo{author}{Raup, D.}, \bibinfo{year}{1992}.
\newblock \bibinfo{title}{Extinction: Bad Genes or Bad Luck?}
\newblock \bibinfo{publisher}{W. W. Norton \& Company, New York}.

\end{thebibliography}

\section{Appendix}

\subsection{Proof of Proposition~\ref{first}}

Recall that $G'$ is a geometric random variable distribution defined as $\min (G,N)$, where $G$ and $N$ are assumed independent, that is
$$
\bP(G' = n) = ((1-p)(1-a))^{n-1}(1-(1-p)(1-a)),\qquad n\ge 1.
$$
and denote by $(A',A'')$ a pair of real random variable defined by
\begin{equation}
\label{eqn:dfn A' A''}
A':=\sum_{i=1}^{G'} A_i \quad \mbox{ and }\quad A'':=\max_{i=1,\ldots, G'}A_i ,
\end{equation}
where $(A_i)$ are i.i.d. copies of $A$, independent of the geometric random variable $G'$. In particular, $A''\le A'$.

First, it is easy to see that 
$$
p_0:=\bP(K=0) = \bE((1-p)^N) = \frac{a(1-p)}{1-(1-a)(1-p)},
$$
and that conditional on $K\not= 0$, $K$ is geometric with success probability $c:=\bP(N\le G)$ (where $N$ and $G$ are assumed independent), that is,
$$
\bP(K= k) = (1-p_0)c (1-c)^{k-1}\quad k\ge 1,
$$
where
$$
c=\bP(N\le G) = \sum_{n\ge 1} \bP(N=n) (1-p)^{n-1}= a\sum_{n\ge 1}(1-a)^{n-1}(1-p)^{n-1} = \frac{a}{1-(1-a)(1-p)}.
$$
Second,  observe that the tree obtained from the coalescent point process after the passage of the field of bullets again satisfies \eqref{eqn : def coal}, and that the coalescence time between two consecutive sampled tips is the maximum of the coalescence times of unsampled tips separating them. In addition, the numbers of unsampled tips between two consecutive sampled tips in the coalescent point process are independent copies of $G'$, independent of $K$, so we get the following joint equality in distribution 
\begin{equation}
\label{eqn:joint}
S_{N}^\star = B\indic{K\not=0}+\sum_{i=1}^{K-1} A_i' + C\quad \mbox{ and }\quad S_{N}^\star(p) = \sum_{i=1}^{K-1} A_i'',
\end{equation}
where the pairs $(A_i', A_i'')$ represent independent copies of $(A',A'')$, independent of $K$ and of $B$ and $C$, which respectively denote the numbers of unsampled tips before the first sampled tip and after the last one. Notice that when $K=0$, $C$ is the only nonzero term. To be specific, $B$ and $C$ are independent, independent of $K$, both distributed as $\sum_{i=1}^{G'-1} A_i$.
In particular, since $\bE(A') = \bE(G') \bE(A)$, we get 
$$
\bE(S_{N}^\star\mid K=k) = \big[\indic{k\not=0} (k \bE(G')-1)+ \bE(G'-1)\big]\ \bE(A) = \big[ (k+1) \bE(G')-\indic{k\not=0}-1\big]\ \bE(A),
$$
so that, thanks to $\bE(K)= p\bE(N)=p/a$ and $\bE(G')= (1-(1-p)(1-a))^{-1}$, we recover
\begin{eqnarray*}
\bE(S_{N}^\star) &= & \big[ (\bE(K)+1) \bE(G')-\bP(K\not=0)-1\big]\ \bE(A) \\
	&=& \left[ \frac{(p/a)+1}{1-(1-p)(1-a)} - \frac{p}{1-(1-p)(1-a)} -1\right]\ \bE(A)\\
	&=& a^{-1}(1-a)\ \bE(A),
\end{eqnarray*}
which is not different from \eqref{eqn:exp Snstar(1)-2}.

Similarly,
\begin{equation}
\bE(S_{N,k}^\star(p)) = (k-1) \bE(A'')\quad \mbox{ and }\quad \bE(S_N^\star(p)) = \bE((K-1)^+) \bE(A''),
\end{equation}
where $x^+$ denotes the positive part of $x$.

Let us compute the two quantities expressed in the right hand side of \eqref{eqn:exp Snstar(p)-1}. First,
$$
\bE((K-1)^+) = (1-p_0)c\sum_{k\ge 1}(k-1)(1-c)^{k-1} = (1-p_0) c^{-1}(1-c) = \frac{p^2a^{-1}(1-a)}{1-(1-a)(1-p)} .
$$ 
Second,
$$
\bP(A''<t)=\bE(\bP(A<t)^{G'}) = F_T(t)\frac{1-(1-a)(1-p)}{1-(1-a)(1-p)F_T(t)},
$$
so that we get
\begin{equation}
\label{eqn:law A''}
\bP(A''> t)= \frac{1-F_T(t)}{1-(1-a)(1-p)F_T(t)},
\end{equation}
and subsequently
\begin{equation}
\bE(A'') = \int_0^T \bP(A''>t)\, dt = \int_0^T\frac{1-F_T(t)}{1-(1-a)(1-p)F_T(t)}\, dt.
\end{equation}

\bigskip

\subsection{Proof of  Proposition~\ref{prop:N=n}}
Each tip is labelled $1,\ldots,n$ from left to right. Let $R_i$ denote the Bernoulli random variable  equal to 1 if tip $i$ is sampled. Set
$$
G_i:=\min\{k\in\{ 1,\ldots ,i-1\}: R_{i-k}=1\},
$$
with the convention that $\min \varnothing= +\infty$, so that $i-G_i$ is the label of the rightmost tip left of $i$ to be sampled. Next, set
$$
B_i:= \indic{R_i=1, G_i <\infty} \max \{A_{i+1-k} : 1\le k \le G_i\}, 
$$
which is the coalescence time between tips $i-G_i$ and $i$ (consecutive tips in the tree spanned by sampled tips).  It is obvious that 
$$
S_n^\star(p) = \sum_{i=2}^n B_i,
$$
and that $\bE(B_i) = p \sum_{j= 1}^i \bE(X_j) \bP(G=j)$, where $G$ is the geometric random variable with parameter $p$ defined in the beginning of this section.  As a consequence,
$$
\bE(S_n^\star(p))= p^2\sum_{i=2}^n\sum_{j=1}^{i-1}(1-p)^{j-1} \bE(X_j) = p^2\sum_{j=1}^{n-1} (n-j)(1-p)^{j-1}\ \bE(X_j).
$$
This expression is satisfying, but we can go further if we want an alternative formulation in terms of the distribution function $F_T$ of the initial node depths. Note that $\bP(X_j<t) = F_T(t)^j$, so that
\begin{eqnarray*}
\bE(S_n^\star(p)) &=& p^2\sum_{i=2}^n\sum_{j=1}^{i-1}(1-p)^{j-1} \int_0^T (1-F_T(t)^j)\, dt\\
	&=& \frac{p^2}{1-p}\int_0^Tdt\,\sum_{i=2}^n\sum_{j=1}^{i-1}(1-p)^{j}(1-F_T(t)^j)\\
	&=& \frac{p^2}{1-p}\int_0^Tdt\,\big(h_n(1-p) - h_n((1-p)F_T(t))\big),
\end{eqnarray*}
where $h_n$ is the function defined by
$$
h_n(x):= \sum_{i=2}^n\sum_{j=1}^{i-1}x^{j} = x^{n}\sum_{j=1}^{n-1}jx^{-j}.
$$
The formula for $h_n$ stems from elementary calculations.\hfill $\Box$

\subsection{Proof of Corollary~\ref{cor}}
\paragraph{Proof.} From Proposition \ref{prop:N=n}, we know that  
$$
\frac{1}{n}\bE(S_n^\star(p)) = \frac{1}{n}\sum_{i=2}^np^2\sum_{j=1}^{i-1}(1-p)^{j-1} \bE(X_j),
$$
which is the Cesaro sum of the sequence with generic term $p^2\sum_{j=1}^{n-1}(1-p)^{j-1} \bE(X_j)$, and so converges to the limit $p^2\sum_{j\ge 1}(1-p)^{j-1} \bE(X_j)$ of this sequence. The convergence of $n^{-1} \bE(S_n(p))$ to this same limit is merely due to the fact that $\bE(S_n(p)) = T+\bE(S_n^\star(p))$. 
\hfill $\Box$

\subsection{Proof of Proposition~\ref{samplingprop}}

The exchangeability of $(C_1, \ldots, C_{k-1})$ is standard.  It is easily seen to hold when conditioning on the leftmost and rightmost sampled species, and obviously holds after integrating over the law of this pair. Of course, the coalescence times are not independent.

As for the formula, first observe that the probability of any configuration of $k$ sampled species among $n$ has the same probability equal to $k! (n-k)!/n!$, so the probability of the event $U_{i,j}$ that the two species with (old) labels $i$ and $i-j$ are sampled and no species among them is sampled, equals 
$$
\frac{k!(n-k)!}{n!}\ \frac{(n-j-1)!}{(k-2)! (n-j-k+1)!} .
$$
On the other hand, summing over all possible such pairs $(i,j)$, we get 
$$
\sum_{u=1}^{k-1} \indic{C_u <t} = \sum_{i=2}^{n}\sum_{1\le j\le \min (i-1, n-k+1)} \indic{U_{i,j}} \indic{\max_{v=i-j+1,\ldots, i} A_v<t}.
$$
Taking expectations and using exchangeability, we get
\begin{eqnarray*}
(k-1) \bP(C_1<t) &=& \frac{k!(n-k)!}{n!} \sum_{i=2}^{n}\sum_{1\le j\le \min (i-1, n-k+1)} \frac{(n-j-1)!}{(k-2)! (n-j-k+1)!} \bP(A<t)^j\\
	&=& k(k-1)\frac{(n-k)!}{n!} \sum_{j=1}^{n-k+1}F_T(t)^j\sum_{i=j+1}^{n} \frac{(n-j-1)!}{(n-j-k+1)!} \\
	&=& k(k-1)\frac{(n-k)!}{n!} \sum_{j=1}^{n-k+1}F_T(t)^j \frac{(n-j)!}{(n-j-k+1)!} 
\end{eqnarray*}
which yields the first formula. For the second formula, see \citet{lam3} or check by hand that for any $x\not=1$,
$$
(1-x)^{-k}\int_x^1 y^{n-k}(y-x)^{k-1}\, dy =\frac{(n-k)!}{n!} \sum_{j=1}^{n-k+1} x^{j-1} \frac{(n-j)!}{(n-j-k+1)!},
$$
by using the change of variables $u= (y-x)/(1-x)$ and expanding the sum.\hfill$\Box$\

\subsection{Proof of Theorem~\ref{thm1}}

We first explain how the random values $(S_n(p);n\ge 1)$ can be simultaneously embedded in the same probability space, in order to be able to speak of almost sure convergence. Consider a sequence $(A_i)_{i\ge 1}$ of independent copies of $A$, and a sequence $(R_i)_{i\ge 0}$ of i.i.d. Bernoulli random variable s with parameter $p$. Consider the coalescent point process built from one edge with tip labelled 0 and length $T$, along with edges with tips labelled $1,2,\ldots$ and respective lengths $A_1, A_2,\ldots$ as in Fig. \ref{fig : treecoal}. The sampled tree is the tree spanned by sampled tips, i.e., tips with labels $i$ such that $R_i=1$. Then we define $S_n(1)$ (resp. $S_n(p)$)  as the PD of the tree (resp. the sampled tree) restricted to tips with labels $i\in\{0,1,\ldots, n-1\}$. This coupling is assumed in the statement of Theorem~\ref{thm1}, which allows us  to obtain almost sure convergences (implying in particular convergence in probability).\\
 
Let $J(i)$ be the label of the $i$-th sampled tip, with the convention that $J(0)=0$. Set
$$
Y_i:=\max_{u=J(i)+1,\ldots,J(i+1)} A_u.
$$
In particular, it is obvious that the random variable $(Y_i)$ represent independent copies of $Y$, so by the law of large numbers, the following convergence holds almost surely 
$$
\lim_{n\to\infty} n^{-1} \sum_{i=1}^n Y_i= \bE(Y) .
$$
Now let $K_n$ the number of sampled tips in $\{0,1,\ldots,n-1\}$. Similarly, the following convergence holds almost surely
$$
\lim_{n\to\infty} n^{-1}K_n = p.
$$
Now for all $n$, $J(K_n)\le n  \le J(K_n+1)$, so similarly as in Eqn. \eqref{eqn:joint}, we can write
$$
\sum_{i=1}^{K_n}Y_i\le S_n^\star(p) \le \sum_{i=0}^{K_n+1} Y_i,
$$
which shows that almost surely $\lim_{n\to\infty}K_n^{-1}S_n^\star(p) = \bE(Y)$, so that almost surely $\lim_{n\to\infty}n^{-1} S_n^\star(p) = p\bE(Y)$. The convergence of $n^{-1} S_n(p)$ follows from the fact that $S_n(p) = S_n^\star(p) +T$.\\

We now turn to the case when the number $k_n$ of sampled species is fixed, and depends on $n$ in such a way that $k_n/n\to p$ as $n\to\infty$.  Let $[x]$ denote the integer part of $x$. Then
$$
S_{n, [n(p-\varepsilon)]}\le S_{n}(p)\indic{p-\varepsilon\le n^{-1}K_n}\quad
\mbox{ and }\quad S_{n}(p)\indic{n^{-1}K_n \le p+\varepsilon} \le S_{n, [n(p+\varepsilon)]},
$$
where the two previous inequalities are stochastic (but could easily be extended to hold almost surely).
As a consequence, we have the following stochastic inequalities between limits in probability
$$
\limsup_{n\to\infty }n^{-1}S_{n, [n(p-\varepsilon)]}\le p\bE(A'') \le \liminf_{n\to\infty}  S_{n, [n(p+\varepsilon)]},
$$
which shows that 
$$
\limsup_{n\to\infty }n^{-1}S_{n, k_n}\le \limsup_{n\to\infty }n^{-1}S_{n, [n(p+\varepsilon)]}\le (p+2\varepsilon) \bE(A''(p+2\varepsilon))
$$
and
$$
\liminf_{n\to\infty }n^{-1}S_{n, k_n}\ge \limsup_{n\to\infty }n^{-1}S_{n, [n(p-\varepsilon)]}\ge (p-2\varepsilon) \bE(A''(p-2\varepsilon)),
$$
where $A''(q)$ is the random variable with the same distribution as $A''$, but after changing $p$ for $q$. Letting $\varepsilon\to 0$ shows that $n^{-1}S_{n, k_n}$ converges in probability to $p\bE(A'')$ (the continuity of the expectation of $A''$ in the parameter $p$ is trivial). The convergence of expectations stems from taking the expectations of the same inequalities.\hfill $\Box$

\subsection{Proof of Theorem~\ref{thm2}}
To keep the dependence on $T$ in mind, we denote by $N_T$ the number of extant species at $T$ and by $S_T(p)$ the remaining PD after the passage of the field of bullets with sampling probability $p$ (which was denoted $S_N(p)$ until now). We will also denote by $A_T$ the typical node depth and by $a_T$ the success probability  of the geometric random variable $N_T$ conditional on $N_T\not=0$. Since rates are not time-dependent, recall that there is a random variable $H$ whose distribution does not depend on $T$ (with tail distribution $\bP(H>t)= (1+(b/r)(e^{rt}-1))$ in the Markovian case), such that $A_T$ is distributed as $H$ conditional on $H<T$ and $a_T=\bP(H>T)$. As a consequence, 
$$
\bE(S_T^\star) =  a_T^{-1}(1-a_T) \ \bE(A_T)=   \bP(H>T)^{-1}\bP(H<T) \ \bE(H \mid H<T) = \frac{1}{\bP(H>T)}\, \bE(H\indic{H<T}),
$$
where, for an event $E$,  ${\bf 1}_{E}$ is the `indicator' function that takes the value 1 when $E$ occurs, and 0 otherwise.
Because $a_T^{-1}=\bE(N_T\mid N_T\not=0) = \bP(H>T)^{-1}$, we see that in the supercritical case, $H$ has an exponential tail and in particular has a finite expectation. In particular,
$$
\lim_{T\to+\infty}a_T\bE(S_T^\star) = \bE(H).
$$
Also thanks to (\ref{eqn:E(A'')}), we have
$$
\bE(S_T^\star(p)) =  \frac{p^2a_T^{-1}(1-a_T)}{1-(1-a_T)(1-p)}\  \bE(A_T''),
$$
where $A_T''$ is the maximum of $G_T'$ independent copies of $A_T$, where $G_T'$ is the geometric random variable with success probability $1-(1-a_T)(1-p)$. Similarly, it is elementary to prove that
$$
\lim_{T\to+\infty}a_T\bE(S_T^\star(p)) = p\bE(Z),
$$
with $Z$ defined as
$
Z:=\max_{i=1,\ldots, G} H_i,
$
where  $(H_i)$ represents independent copies of $H$, independent of the geometric random variable $G$ with success probability $p$. 

Last, since $S_T^\star(p) = S_T(p)-T$, we get 
$$
\lim_{T\to+\infty}a_T\bE(S_T(1)) = \bE(H)\quad\mbox{ and }\quad \lim_{T\to+\infty}a_T\bE(S_T(p)) = p\bE(Z),
$$
and subsequently
$$
\lim_{T\to+\infty}\frac{\bE(S_T(p))}{\bE(S_T(1))} = \frac{p\bE(Z)}{\bE(H)}.
$$

Now,  we can embed all trees with stem age $T$ in the same probability space. Consider a sequence $(H_i)_{i\ge 1}$ of independent copies of $H$, and  build a tree starting with one infinite half-line with tip labelled 0 and edges with tips labelled $1,2,\ldots$ and lengths $H_1, H_2,\ldots$, as on Fig.~\ref{fig : treecoal}. Build the sample tree as in the previous subsection thanks to a sequence of i.i.d. Bernoulli random variable with parameter $p$. Define $N_T:=\min\{i\ge 1: H_i >T\}$. Then we define $S_T(1)$ (resp. $S_T(p)$)  as the PD of the tree (resp. the sampled tree) restricted to the tips labels $\{0,1,\ldots, N_T-1\}$. Regardless of edge lengths, we have exactly the same picture as in the previous subsection after replacing $A$ with $H$. In this setting, Theorem \ref{thm1} yields the almost sure convergence of $S_n(p)/S_n(1)$ to $p\bE(Z)/\bE(H)$. This almost sure convergence ensures the almost sure convergence of $S_n(p)/S_n(1)$ along the subsequence $\{N_T:T>0\}$, which is exactly the almost sure convergence of $S_T(p)/S_T(1)$. This establishes the claim in Theorem~\ref{thm2}.

\subsection{Fluctuations around the deterministic almost sure limit}
\label{subsec:clt}
Recall $G$, $Y$ and $Y'$ from main text. 
Recall that $\bE(Y)$ has been computed in \eqref{eqn:EY} and that $\bE(G) = p^{-1}$, so we have
$$
\bE(Y') = \bE(G) \,\bE(A) = p^{-1}\,\bE(A).
$$
As a consequence, equation \eqref{eqn:main1} can be read as the  convergence (a.s. or in probability) of $S_n(p)/S_n(1)$, as $n\to\infty$, to
$$
\pi_T(p) = \frac{p\bE(Y)}{\bE(A)} = \frac{\bE(Y)}{\bE(Y')} .
$$
Actually, this is not surprising since 
$$
\frac{S_{J(n)}(p)}{S_{J(n)}(1)} = \frac{\sum_{i=1}^nY_i}{\sum_{i=1}^nY_i'} ,
$$
where $J(n)$ is the label of the $n$-th sampled species, and the pairs $(Y_i, Y_i')$ represent independent copies of the pair $(Y, Y')$. Then the strong law of large numbers ensures the a.s. convergence of $S_{J(n)}(p)/S_{J(n)}(1)$, as $n\to\infty$, to $\bE(Y)/\bE(Y')$. Forgetting momentarily the fact that $(S_{J(n)}(p)/S_{J(n)}(1))$ is only a subsequence of $(S_n(p)/S_n(1))$, this is the aforementioned result. With this new presentation, we can go further and apply the central limit theorem to get, by an elementary Taylor expansion, the following convergence in distribution
$$
\lim_{n\to\infty}\sqrt{n}\left(\frac{S_{J(n)}(p)}{S_{J(n)}(1)} -  \frac{\bE(Y)}{\bE(Y')}\right) = V,
$$
where $V$ is a centered, Gaussian random variable with variance $\sigma^2$ given by \eqref{sigma}.
Using the fact that $J(n)$ is the sum of $n$ independent (geometric) random variables with expectation $1/p$, we get that $J(n)/n$ goes to $1/p$ almost surely. We can then write
$$
\lim_{n\to\infty}\sqrt{J(n)}\left(\frac{S_{J(n)}(p)}{S_{J(n)}(1)} -  \frac{\bE(Y)}{\bE(Y')}\right) = p^{-1/2}V,
$$
Now it is easy to see that the difference between $S_{J(n)}(p)/S_{J(n)}(1)$ and $S_{k}(p)/S_{k}(1)$ for $J(n)\le k < J(n+1)$ is of the order of $1/n$, so we get the convergence in distribution of the whole sequence, and not only of the subsequence indexed by $J(n)$.

Since we have already displayed formulae for $\bE(Y)$ and $\bE(Y')$, it is sufficient, to evaluate $\sigma$, to have the values of $\bE(Y^2)$, of $\bE(Y'^2)$ and of $\bE(YY')$.
First, note that 
$$
\bE(A^2) = \int_0^T 2t\,(1-F_T(t))\, dt,
$$
and similarly
$$
\bE(Y^2) = \int_0^T \frac{2t\,(1-F_T(t))}{1-(1-p)F_T(t)}\, dt.
$$
Elementary calculations provide the following formula
$$
\bE(Y'^2) = p^{-1}\bE(A^2) + 2p^{-2}(1-p) \bE(A)^2.
$$
The last expression is given in the following statement.
\begin{lem}
\label{last one}
We have
$$
\bE(YY') = Tp^{-1}\bE(A) - \int_0^T \frac{pg(t)}{(1-(1-p)F_T(t))^2}\, dt,
$$
where $g(t) = \bE(A\indic{A<t})=tF_T(t) - \int_0^t F_T(u)\, du$.
\end{lem}

\end{document}